\documentclass[10pt,aps,prb,twocolumn,showpacs,superscriptaddress,citeautoscript,longbibliography,floatfix]{revtex4-2}
\usepackage{graphicx} 
\usepackage{mathtools}
\usepackage{amsmath} 
\usepackage{tensor}
\usepackage{cancel}
\usepackage{leftindex}
\usepackage{nccmath}
\usepackage{xfrac} 
\usepackage{bm} 
\usepackage{natbib}
\usepackage{natmove}
\usepackage{color} 
\usepackage{dcolumn} 
\newcolumntype{.}{D{.}{.}{2.1}}
\newcolumntype{-}{D{.}{.}{4.0}}
\usepackage{makecell}
\usepackage{multirow}
\usepackage{braket}
\usepackage{xcolor}

\renewcommand{\today}{\number\day \space \ifcase \month \or January\or February\or March\or April\or May \or June\or July\or August\or September\or October\or November\or December\fi \space \number\year} 
\def\m1r{\multicolumn{1}{r}}

\usepackage[colorlinks=true, linkcolor=blue, citecolor=blue, urlcolor=blue, linktoc=page, bookmarks=false, pdfstartview={FitH}, pdfborder={0 0 0.0 [3 3]}]{hyperref} 
\usepackage{cleveref} 
\crefname{figure}{Fig.}{Figs}
\crefname{table}{Table}{Tables}

\begin{document}

\title{Smooth Overlap of Spin Orientations: Machine Learning Exchange Fields for Ab-initio Spin Dynamics}
\author{Yuqiang Gao}
\email[Email: ]{y.gao@ahnu.edu.cn}
\affiliation{School of Physics and Electronic Information, Anhui Province Key Laboratory for Control and Applications of Optoelectronic Information Materials, Anhui Normal University, Wuhu 241000,  PR China}
\affiliation{Faculty of Science and Technology and MESA$^+$ Institute for Nanotechnology,
University of Twente, P.O. Box 217, 7500 AE Enschede, The Netherlands}
\author{Menno Bokdam}
\email[Email: ]{M.Bokdam@utwente.nl}
\affiliation{Faculty of Science and Technology and MESA$^+$ Institute for Nanotechnology,
University of Twente, P.O. Box 217, 7500 AE Enschede, The Netherlands}
\author{Paul J. Kelly\thanks{corresponding author}}
\email[Corresponding author: ]{P.J.Kelly@utwente.nl}
\affiliation{Faculty of Science and Technology and MESA$^+$ Institute for Nanotechnology,
University of Twente, P.O. Box 217, 7500 AE Enschede, The Netherlands}
\date{\today}
\begin{abstract}
Ab-initio molecular dynamics (AIMD) refers to the solution of Newton's equations of motion for ions with forces ${\bf f}_i = -\partial E / \partial {\bf R}_i$ calculated from self-consistent electronic structure calculations. So-called machine-learning force field (ML-FF) schemes parameterize the potential energy surface very efficiently and make it possible to extend the time scale of AIMD simulations by orders of magnitude. 
The Landau-Lifshitz equation describes the dynamics of atomic magnetic moments ${\bf m}_i$ in effective fields ${\bf h}_i = -\partial E / \partial {\bf m}_i$ which in addition to containing external magnetic fields, describe contributions from interatomic exchange interactions, long-range dipolar interactions, anisotropy fields etc. 
In this publication, we add the magnetic degrees of freedom to the widely used Gaussian Approximation Potential of machine learning and present a model that describes the potential energy surface of a crystal based on atomic coordinates and noncollinear magnetic moments. 
Incorporating the translational, rotational, inversion and permutational symmetries of magnetic interactions, the ML model can describe various magnetic interactions expanded into two-body, three-body terms, etc., in  the spirit of the atomic cluster expansion.
Assuming an adiabatic approximation for the spin directions, the ML model depends solely on the positions and orientations of atomic spins and is computationally efficient enough to make  coupled ab initio molecular and spin dynamics possible. 
To illustrate the ML model, we implement a two-body form for the interatomic exchange interaction. 
Comparing the total energies and local exchange fields predicted by the model for noncollinear spin arrangements with the results of constrained noncollinear density functional calculations for bcc Fe yields very good results, with agreement on the level of 1 meV/spin for the total energy. 
\end{abstract}

\maketitle
\section{Introduction}
\label{sec:intro}

At finite temperatures, the magnetic moments of magnetic materials fluctuate in magnitude and direction influencing the potential energy surface (PES) that describes the thermal motion of the ions.  
The purpose of this publication is to extend to itinerant magnetic materials recently developed methods for efficiently parameterizing the PES to perform first-principles (FP) molecular dynamics (MD) simulations. 
This requires taking the spin degrees of freedom into consideration by extending the concept of ``machine-learning force fields'' (ML-FF) to ``machine-learning exchange fields'' (ML-EF) with a view to performing coupled ionic and spin dynamics. 
We begin with a brief summary of recent relevant progress in first-principles (or {\it ab-initio}) molecular dynamics, FP-MD (or AIMD).

In the Car-Parrinello AIMD scheme \cite{Car:prl85b}, the interactions between the $N$ atoms comprising a molecule or (the periodic supercell of) a solid are described in terms of full quantum mechanical (QM) solutions for the constituent electrons. 
AIMD is ultimately limited by the need to describe the electron and ion dynamics on the same short time scale necessary to keep the electrons in the ground state \cite{Behler:prl07} required by density functional theory (DFT) \cite{Hohenberg:pr64}. 
For metals with no gap between occupied and unoccupied states, it turns out to be more efficient to combine long molecular-dynamics (MD) time steps for the ions with full self-consistent field (SCF) solutions of the Kohn-Sham equations of DFT for the electrons \cite{Kohn:pr65}. 
Every such step yields the total electronic energy that plays the role of a potential energy for the ionic motion, the forces acting on all the atoms, and the stress tensor \cite{Kresse:prb99, Torrent:cms08}. 
The resulting Born-Oppenheimer MD (BO-MD) procedure is limited by the DFT calculation to time scales of tens of picoseconds (ps) for thousands of atoms \cite{Behler:prl07} (hundreds of ps for hundreds of atoms \cite{Behler:jcp16}). 
For comparison, typical optical vibration frequencies are in the range $10^{12}-10^{13}$ Hertz.

In these simulations, many of the computationally expensive DFT calculations are unnecessary because at temperatures at which bonds are only seldom broken, much of the simulation time is spent repeatedly exploring a tiny portion of the 3$N$ dimensional coordinate space. 
The effectiveness of modelling a huge diversity of systems using periodic boundary conditions is  such that $N$ is determined by the size of the supercell required to model the system of interest and far smaller than the $N\approx 10^{23}$ atoms composing a typical solid. 
Typically $N$ should be of order $10^3$ to avoid artifacts of the artificial periodicity but such large systems are usually prohibitively expensive. 
For metallic or nonpolar materials, the range of the interatomic force constants is quite limited, to of order $\sim 30$ atoms (the long range electrostatic interactions in polar solids can be taken care of classically without significantly increasing the computational cost). 
The challenge is to find an efficient parameterization of the DFT BO potential energy surface (PES) in this enormously reduced but still very large coordinate space.

\subsection{Machine Learning approaches for the PES}
\label{ssec:ml}

Behler and Parrinello decomposed the total potential energy $U_{\rm tot}$ of a system of interacting atoms into a sum of local atomic contributions $U_i$ 
\begin{equation}
    U_{\rm tot}=\sum_{i=1}^N U_i(\{ {\bf r}_j \} - {\bf r}_i),
\label{Eq:PES}
\end{equation}
reducing the problem of describing the energetics of an infinite crystal to one of describing the energetics of a large cluster and then replaced the atomic Cartesian coordinates $\{ {\bf r}_j \}$ on which $U_i$ depends with coordinates (``descriptors'') that reflect the rotational, reflectional, inversion and permutational symmetry of atom $i$, so-called ``atom-centered symmetry functions'' (ACSF) \cite{Behler:prl07, Behler:jcp11}.
These innovations received a great deal of attention and stimulated much subsequent activity.  
Observing that the PES is a relatively smooth function of the atomic coordinates on which it depends and using the same atomic decomposition \eqref{Eq:PES} of the potential energy, Bartok {\it et al.} introduced the density $\rho_i({\bf r})$ at position ${\bf r}$ of the atomic neighbours of each atom $i$ 
\begin{equation}
\rho_i({\bf r}) = \sum_j^{r_{ij} < r_{\rm cut}}  f_{\rm cut}(r_{ij}) \,\, \delta({\bf r}-{\bf r}_{ij}),
\label{Eq:rho}
\end{equation}
in terms of $\delta$ functions at each atomic position \cite{Bartok:prl10}. 
Here the index $j$ runs over the neighbors of atom $i$ at ${\bf r}_i$ within some radius $r_{\rm cut}$, ${\bf r}_{ij} = {\bf r}_j - {\bf r}_i$ and $r=|{\bf r}|$. 
The smooth cutoff function $f_{\rm cut}(r)$ removes information about the structure beyond $r_{\rm cut}$ and a commonly made choice is $f_{\rm cut}(r)= [1+\cos(\pi r/r_{\rm cut})]/2$ \cite{Behler:prl07, Bartok:prl10, Jinnouchi:prb19}.
Ideally, $\rho({\bf r})$ should characterize an atomic environment uniquely. 
Bartok {\it et al.} described the radial dependence in terms of an angle $\theta_0 = r/r_0$, 
expanded $\rho$ in ``hyperspherical'' harmonics and reduced the determination of the BO-PES to interpolating 
atomic energies $U_i$ in this space. 
To do this, they introduced a nonparametric method called ``Gaussian process regression'' and termed the resulting PES the ``Gaussian Approximation Potential'' (GAP) \cite{Bartok:prl10}.  
Because of the delta functions used to describe the atomic positions, the expression \eqref{Eq:rho} for $\rho_i({\bf r})$ is not smooth and this leads to difficulties characterizing the similarity of different atomic environments. 
Replacing the $\delta$ functions with Gaussians led to a smooth measure of similarity that was termed ``Smooth Overlap of Atomic Positions'' (SOAP) \cite{Bartok:prb13}.   

In spite of the great reduction in the size of the coordinate space made possible by the approximations outlined above, a huge number of calculations is still needed to determine the BO-PES to perform AIMD, even for simple materials. 
A considerable improvement was described by Jinnouchi {\it et al.} whereby new DFT calculations are only performed for an uncharted volume of coordinate space when it is explored by the MD simulation, so-called ``on-the-fly machine learning (ML)''. 
The criterion for deciding to perform a new DFT calculation is in essence based upon the distance of the unexplored region of coordinate space from regions already explored and the estimate of the error in the potential energy and forces is based upon a Bayesian regression analysis that lends itself to full automation \cite{Jinnouchi:prl19, Jinnouchi:prb19}.

{\subsection{Including magnetism in ML-FFs}
\label{ssec:imag}

In the ``machine-learning force-field (ML-FF)'' approach just sketched, the electronic degrees of freedom have been effectively integrated out and Newton's equations of motion are solved for the ionic degrees of freedom using an effective force field that is determined quantum mechanically and interpolated in the coordinate space. 
As a result, a much longer time step appropriate to the ionic dynamics can be used. 
With a view to treating lattice and spin dynamics on time scales corresponding to phonons and magnons and not electrons, our aim is to develop an analogous procedure for magnetic materials.
To do so, we will integrate out the spatial distribution of the spin density to yield atomic moments ${\bf m}_i$ and effective (``exchange'') fields ${\bf h}_i$ which enter Landau-Lifshitz-like (LL) equations that are used to describe the dynamics of atomic magnetic moments in what is called ``Atomistic Spin Dynamics (ASD)'' \cite{Skubic:jpcm08, Eriksson:17}.

The LL equations describe the time variation of a continuous magnetization ${\bf M}({\bf r},t)$ in an effective magnetic field ${\bf H}_{\rm eff}(t)$ as 
\begin{equation}
 \frac{\partial{\bf M}({\bf r},t)}{\partial t} = - \gamma {\bf M}({\bf r},t) \times {\bf H}_{\rm eff}({\bf r},t)
\label{Eq:LL}
\end{equation}
where $\gamma$ is the gyromagnetic ratio $g\mu_{B}/\hbar$, $\mu_B=e\hbar/2 m_e$ is the Bohr magneton, $g\sim 2$ and 
\begin{equation}
   {\bf H}_{\rm eff}({\bf r},t) = -\frac{\partial F[{\bf M]}}{\partial {\bf M}}
\label{Eq:Heff}
\end{equation}
in terms of the free energy $F[{\bf M]}$ \cite{Miltat:02, Tserkovnyak:rmp05}.
The effective field can be decomposed into contributions from applied, dipolar demagnetization, crystal-anisotropy and exchange fields according to the corresponding contributions to $F[{\bf M]}$. 

In this manuscript we will be focussing on bulk itinerant ferromagnets like Fe so that the applied, demagnetization and anisotropy fields can be neglected by comparison with the interatomic exchange field. 
In \eqref{Eq:LL} and \eqref{Eq:Heff}, ${\bf M}({\bf r},t)$ and ${\bf H}_{\rm eff}({\bf r},t)$ are vector fields and solution of the LL equation forms the subject of ``micromagnetism'' \cite{Miltat:02}.
Since we are interested in developing a first-principles description of magnetization dynamics, we will replace the continuum vector fields with discrete atomic quantities \cite{Atxitia:jpd17} as in ASD \cite{Skubic:jpcm08, Eriksson:17}. 
We will present a descriptor for noncollinear spins which takes into account both the spin orientation (the ``Smooth Overlap of Spin Orientations'' analogous to SOAP) and atomic coordinates within the GAP framework \cite{Bartok:prl10, Bartok:ijqc15}. 
Implementation of this ML model for interatomic exchange shows that it captures the DFT total energies and effective (exchange-correlation) field of non-collinear spin structures at different temperatures demonstrating its promise. 

The ACSF \cite{Behler:prl07, Behler:jcp11, Behler:cr21} and GAP-SOAP \cite{Bartok:prl10, Bartok:prb13, Deringer:cr21} descriptors as well as the ``spectral neighbor analysis potential'' (SNAP) \cite{Thompson:jcompp15} and ``moment tensor potential'' (MTP) \cite{Shapeev:mms16} approaches can be considered to be special cases in a systematic body-ordered expansion of the energy called the atomic cluster expansion (ACE) \cite{Drautz:prb19}. 
All have been extended to consider spin degrees of freedom  \cite{Drautz:prb20, Eckhoff:npjcm21, ZhangP:prl21, *ZhangP:npjcm23, Domina:prb22, Chapman:sr22, Novikov:npjcm22, *Kotykhov:cms24, YuH:prb22, *YuH:prb24, Li:ncs23, Suzuki:prb23} and frequently tested using Heisenberg Hamiltonians. 
Very recently the variant of ACE generalized to include spin degrees of freedom \cite{Drautz:prb20} was trained with a very extensive database of spin-density functional calculations to describe the magnetic properties of iron \cite{Rinaldi:npjcm24}.
This will be discussed in more detail in \Cref{Sec:Discussion}.  

\subsection{Itinerant ferromagnets}
\label{ssec:ifm}

Before discussing the SOSO (Smooth Overlap of Spin Orientations) descriptor, it will be useful to consider the energetics of various spin interactions in order to motivate our choice for its form. 
For the open shell atoms Fe, Co and Ni, that in crystalline form exhibit ferromagnetism (FM), Hund's first rule describes how the unpaired valence electrons minimize their mutual Coulomb repulsion by maximizing the alignment of their spins. 
The first row of \Cref{TableI} shows the Hund's rule energy gain for free atoms in the local spin density approximation (LSDA) compared to the local density approximation (LDA) assuming a $3d^{n-2}4s^2$ atomic configuration \cite{Kotochigova:pra97, *Kotochigova:pra97e}.
The spin-polarization energy $E_{\rm SP}$ is approximately $\frac{1}{4} m^2 I_{\rm xc}$ in terms of the magnetic moments of 4, 3 and 2 $\mu_B$ for Fe, Co and Ni, respectively and the Stoner parameter $I_{\rm xc}$ that has a value of 0.9--1.0~eV for the late 3$d$ elements \cite{Gunnarsson:jpf76, Poulsen:jpf76, Janak:prb77}. 

The $3d^{n-2} 4s^2$ electronic configuration for a neutral first-row transition-metal atom becomes $3d^{n-1} 4s^1$ in the metal solid with a corresponding reduction of the magnetic moment by $1 \mu_B$.
Incomplete spin-polarization of the energy bands leads to even smaller ferromagnetic moments of $\sim$ 2.2, 1.6 and 0.6 $\mu_B$ for Fe, Co and Ni, respectively (fifth row of \Cref{TableI}) for which the expected spin-polarization energy is $\sim$ 1.2, 0.6 and 0.1 eV/atom.
This is substantially larger than the energy difference $E_{\rm NM}-E_{\rm FM}$ calculated for a ferromagnetically ordered ground state with magnetic moment $m=n_{\uparrow}-n_{\downarrow}$ and energy $E_{\rm FM}$ and a spin-compensated or nonmagnetic (NM) state with $n_{\uparrow}=n_{\downarrow}$ and energy $E_{\rm NM}$ shown in the third row of the Table \cite{footnote4}. 
This comes about because the promotion of $m/2$ electrons from down-spin to up-spin states entails a loss of kinetic energy \cite{Kubler:00, Mohn:03}.
An additional complicating factor is that $E_{\rm FM}$ also contains contributions from {\it interatomic} exchange interactions between (electrons on) neighbouring atoms usually described in terms of the effective Heisenberg Hamiltonian for classical moments ${\bf m}$
\begin{equation}
\label{Eq:SHEX}
E=-\frac{1}{2}  \sum_{i \neq j} J_{ij} {\bf m}_i \cdot {\bf m}_j 
\end{equation}
where $J_{ij}$ describes the exchange interaction between atoms $i$ and $j$ with fixed magnetic moments ${\bf m}_i$ and ${\bf m}_j$, respectively. 

\begin{table} [t]
\caption{Some characteristic magnetic energies (in meV/atom unless stated otherwise) for Fe, Co and Ni at their experimental lattice constants. 
Energies $E_{\rm NM}$, $E_{\rm FM}$ and $E_{\rm AFM}$ are for nonmagnetic (NM), ferromagnetic (FM) and antiferromagnetic (AFM) states, respectively.
$E_{\rm dd}$ is the magnetostatic dipole-dipole energy of the corresponding magnetic moments at nearest-neighbour separations.
$E_{\rm MAE}$: magnetocrystalline anisotropy energy ($\mu$eV/atom).
}
\begin{ruledtabular}
\begin{tabular}{lccc}
                                & Fe (bcc)        & Co (hcp)        & Ni (fcc) \\                                                                   
\hline
$E_{\rm SP}=E_{\rm LDA}-E_{\rm LSDA}$ & 3540$^a$  & 2070$^a$        & 950$^a$ \\
$m_{\rm atomic}(\mu_B)$         & 4               & 3               & 2   \\   
$E_{\rm NM} -E_{\rm FM}$        & 561             & 208             & 58  \\
$E_{\rm AFM}-E_{\rm FM}$        & 459             & 160             & 45  \\
$m_{\rm FM};m_{\rm AFM}(\mu_B)$ & 2.26; 1.63      & 1.57; 0.96      & 0.62; 0.17 \\     
$\hbar \omega_{\rm magn}(k)$    & $\sim 400^b$(H) & $\sim 570^b$(M) & $\sim 350^b$(X) \\
$\hbar \omega_{\rm phon}(k)$    & $\sim 38^c$(H)  & $\sim 36^d$(M)  & $\sim 37^c$(X) \\
$E_{\rm dd}$ ($\mu$eV)          &  34.9           & 20.3            & 2.5 \\
$E_{\rm MAE}$ ($\mu$eV)         &  1.4$^e$        & 65$^f$          & 2.7$^e$ \\
$a_{\rm expt}$(\AA)             & 2.86$^h$        & $a=b=2.507^i$   & 3.524$^j$  \\
                                &                 & $c=4.069^i$     &   \\
\end{tabular}
\end{ruledtabular}
$^a$Ref.\cite{Kotochigova:pra97}
$^b$Ref.\cite{Halilov:prb98b} 
$^c$Ref.\cite{DalCorso:prb00} 
$^d$Ref.\cite{Wakabayashi:prb82}             
$^e$Ref.\cite{Escudier:adp75}  
$^f$Ref.\cite{Paige:jmmm84}    
$^h$Ref.\cite{Basinski:prsla55}
$^i$Ref.\cite{Vincent:crh67}
$^j$Ref.\cite{Taylor:jim50}
\label{TableI}
\end{table}

A measure of the interatomic exchange interactions is the energy difference $E_{\rm AFM}-E_{\rm FM}$ where $E_{\rm AFM}$ corresponds to the electronic ground state energy when the magnetic moments of neighbouring atoms are constrained to be aligned antiparallel; for bcc Fe, hcp Co and fcc Ni, these are, respectively, the nearest-neighbour atoms along the [111], [100] and [110] directions. 
These interatomic exchange energies are reflected in the optical magnon energies $\hbar \omega_{\rm magnon}$ shown in the sixth row of \Cref{TableI} and seen to be sizeable. 
Their large values compared to $kT$ suggests that optical-magnon like excitations and Stoner excitations (whereby discrete spin flips occur on individual atoms) will be energetically unfavourable at elevated temperatures. 
Indeed, it is well known that long wavelength spin waves based upon \eqref{Eq:SHEX} with energies $\hbar \omega_{\rm magnon}(q)$ dominate the behaviour of magnetic materials at finite temperatures. In the absence of magnetocrystalline anisotropy, $\hbar \omega_{\rm magnon}(q) \rightarrow 0$ in the limit $q \rightarrow 0$.  

A central assumption of \eqref{Eq:SHEX} is that the size of the atomic magnetic moments is fixed. 
This assumption is not true in general for itinerant magnets where the fifth row of \Cref{TableI} shows a very substantial quenching of the magnetic moments in the AFM structures. 
For Fe, the magnetic moment of 2.2$\, \mu_{\rm B}$ per atom for FM ordering decreases to a value of about 1.6$\, \mu_{\rm B}$ for AFM ordering. 
For Co and Ni the intraatomic exchange is too weak to stabilize the magnetic moments which are almost completely quenched and $E_{\rm AFM} \approx E_{\rm NM}$. 
$E_{\rm NM}$, $E_{\rm FM}$ and $E_{\rm AFM}$ as well as $m_{\rm FM}$ and $m_{\rm AFM}$ are plotted in \Cref{fig:GGA_EMa} as a function of the lattice parameter for bcc Fe \cite{footnote4} which shows that even the modest 4\% lattice expansion between $T=0\,$K and the Curie temperature $T_{\rm C}=1043\,$K leads to a $\sim 5\%$ increase of the atomic moment because the larger lattice constant leads to a reduced bandwidth of the $d$ bands; for the AFM state the change is much larger.  
We will see that this variation of the magnitude of the magnetic moments does not need to be described explicitly.

\begin{figure}[t]
\includegraphics[width=8.6cm]{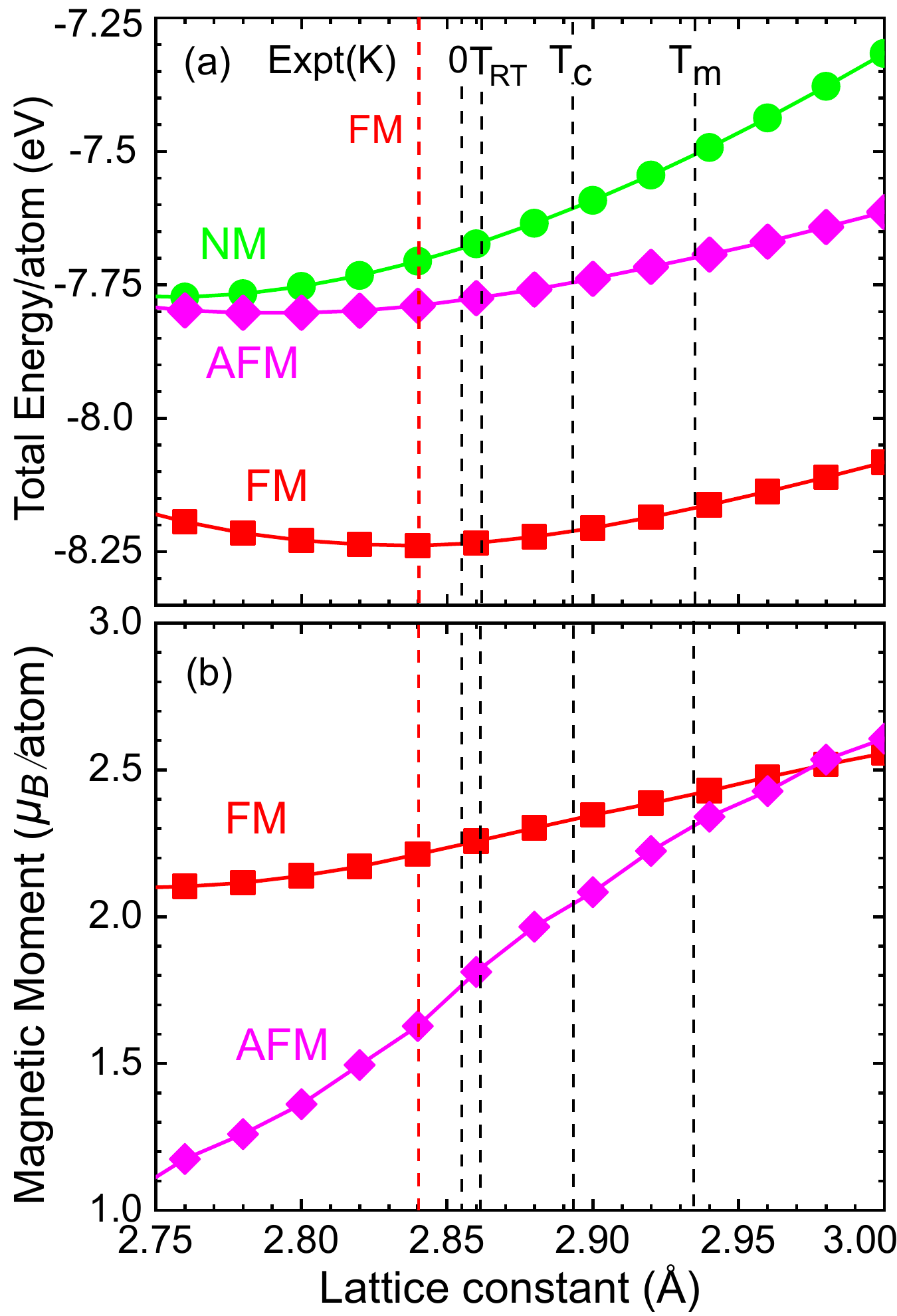}
\caption{\label{fig:GGA_EMa} (a) Total energy per atom as a function of the lattice constant $a$ for nonmagnetic (NM), ferromagnetic (FM) and antiferromagnetic (AFM) ordering of the magnetic moments in bcc Fe in the GGA \cite{footnote4}. The black vertical dashed lines indicate experimental lattice constants at $T=0\,$K, room temperature (RT; $T=293\,$K), the Curie temperature $T_{\rm C}=1043\,$K and the melting temperature $T_m=1807\,$K \cite{Basinski:prsla55} while the red vertical dashed line indicates the theoretical (GGA) equilibrium lattice constant for FM Fe at $T=0\,$K.
(b) Size of the atomic magnetic moment of bcc Fe as a function of the lattice constant $a$ for FM and AFM states calculated in the GGA.
}
\end{figure}

Row seven of \Cref{TableI} shows the energy $\hbar \omega_{\rm phonon}$ of the optical phonon modes of Fe (at the H point), Co (M point) and Ni (X point) \cite{DalCorso:prb00, Wakabayashi:prb82}. 
These are seen to be roughly an order of magnitude smaller than the corresponding $\hbar \omega_{\rm magnon}$ energies. 
For example, for hcp Co, the optical frequencies for phonons and magnons at the M point are 8.6 THz ($\sim$36~meV) \cite{Wakabayashi:prb82} and 138 THz ($\sim 570\,$ meV) \cite{Durhuus:jpcm23}, respectively. 
As the temperature of a ferromagnet is increased, the amplitude of the lattice vibrations increases leading to variation of the separation between atoms and correspondingly, of the exchange interaction  $J_{ij}(r)$ between them.
At the same time, thermal lattice expansion will cause the average separation between atoms to increase. 
This makes it important to be able to perform molecular and spin dynamics simultaneously in order to be able to study the effect of phonon-magnon coupling. 

Row nine of \Cref{TableI} details the magnetocrystalline anisotropy energy which is seen to be orders of magnitude smaller than other characteristic energies for cubic materials and can be safely neglected in this publication which will focus on bcc Fe.
The Dzyaloshinskii-Moriya interaction (DMI) energy vanishes for structures with inversion symmetry and will also not be considered further here \cite{Camley:ssr23}; for a variety of interfaces it has a value of order $10^{-4}\,$eV \cite{Chen:prl13, *Yang:natm18, *Chen:natc13}.  
Both interactions require spin-orbit coupling in the electronic structure calculations making their inclusion computationally very demanding.

\subsubsection*{Smooth Overlap of Spin Orientations}
\label{sssec:soso}

The main challenge in developing an effective ML model for magnetic materials is to efficiently represent the local environment seen by the magnetic moments ${\bf m}_i$ on atoms $i$ characterized by their magnitudes $m_i$, orientations ${\bf e}_i={\bf m}_i/m_i$ and positions ${\bf r}_i$. 
At finite temperatures these are subject to longitudinal and transverse fluctuations \cite{Heine:prb21} that lead to a huge phase space for the non-collinear magnetic system. 
Because typical interatomic exchange parameters $J_{ij}$ are small (e.g. $< 30$ meV for Fe, Co, Ni \cite{Pajda:prb01}) compared to the characteristic electronic energies such as intraatomic exchange, interatomic hopping, etc. the moments reorient slowly on a time scale of $\sim \,$ps while their magnitudes are determined by the change of the electronic wave functions and vary rapidly on time scales of $\sim \,$fs. 
By analogy with the Born-Oppenheimer approximation in molecular dynamics \cite{Born:adp27}, an adiabatic approximation between the rotation of the moment orientation and fluctuations of the magnitude of the moment can be made \cite{Antropov:prl95, *Antropov:prb96}. 
We will see that variation of the magnitudes $\{m_i\}$ of the magnetic moments can be included implicitly so that a ML model can be constructed solely based on the moment orientation and its location (which is just the corresponding atomic position) thereby reducing the phase space for the non-collinear system and improving the efficiency of the ML model. 

An important contribution to the success of ML-FFs was the SOAP (smooth overlap of atomic positions) whereby the delta functions used to describe atomic positions in \eqref{Eq:rho} were replaced with Gaussians yielding a smooth measure of the similarity of two atomic structures. 
Describing spin orientations analogously with Gaussians yields a smooth measure of the similarity of two spin structures in what we call the ``Smooth Overlap of Spin Orientations'' (SOSO).  

The paper is organized as follows. \Cref{sec:method} introduces the methodology used to build the descriptor for the non-collinear spin structures. 
\Cref{ssec:SOAP} briefly recapitulates the power spectrum of the SOAP descriptor. 
and the SOSO analogue for spins is introduced in \Cref{ssec:SOSO}.
We derive a two-body descriptor and kernel for magnetic exchange in \Cref{ssec:exchange} and test its performance in \Cref{Sec:train-test}. 
Because the energies of the magnetic dipole-dipole, magnetic anisotropy (and other) interactions are so small in cubic Fe compared to the exchange interaction, we will only consider the latter here. 
In \Cref{ssec:Test-Heisenberg}, the ML model is tested on a Heisenberg Hamiltonian with varying magnitude of the magnetic moment and atomic positions. 
In \Cref{ssec:Test-DFT}, we perform constrained self-consistent first-principles calculations for non-collinear spin systems to train and test the ML model. 
We compare our model to recent related work in \Cref{Sec:Discussion} before summarizing in  
\Cref{Sec:Conclusion}. 
We derive descriptors and kernels for interatomic exchange (three-body) in \Cref{Appendix:A} and for the dipole-dipole interaction in \Cref{Appendix:B}; their implementation will be the subject of a future publication.  

\section{METHODOLOGY }
\label{sec:method}

In the spirit of the cluster expansion \cite{Kikuchi:pr51}, the total energy can be expanded in a sum of $n$-body (cluster) interaction terms ($n=1,2,... ,\infty$) and a key to the success of ML-FFs was the incorporation of the translational, rotational, inversion and permutational invariance in the formulation of the atomic interactions \cite{Behler:prl07, Behler:jcp11, Bartok:prl10, Bartok:prb13, Jinnouchi:prl19, Jinnouchi:prb19}.
Although there have been discussions about its completeness, it appears that the accuracy of present ML force fields containing two- and three-body terms is already sufficient for many practical applications \cite{Jinnouchi:prb19,Drautz:prb19, Maresca:npjcm18, Davidson:prm20, Vandermause:npjcm20, Jansen:prb23, Lahnsteiner:prb22}. 
It has been argued that the decomposition of the total energy into a sum of atom-centered contributions mitigates the influence of fundamental deficiencies of this approach \cite{Pozdnyakov:prl20}.

We expect that mapping the local spin configurations onto symmetry-adapted descriptors will greatly reduce the phase space of noncollinear spin systems needed to describe the magnetic PES and consider this qualitatively using the phenomenonlogical ``spin'' Hamiltonian \cite{footnote3, White:83} 
\begin{equation}
\begin{aligned}
\label{Eq:SH}
H_{\rm spin} & =  -\frac{1}{2}  \sum_{\langle i,j \rangle} J_{ij}(r_{ij}) {\bf m}_i \cdot {\bf m}_j \\
  & - \frac{\mu_0}{4\pi} \sum_{\langle i,j \rangle} 
\frac{ 3({\bf m}_i \cdot {\bf r}_{ij})({\bf m}_j \cdot {\bf r}_{ij})-{\bf m}_i \cdot {\bf m}_j } 
{|r_{ij}|^3}  \\
& - K \sum_i ({\bf m}_i \cdot {\bf e}_K)^2 - \sum_{\langle i,j \rangle} D_{ij} \cdot [{\bf m}_i \times {\bf m}_j ]
\end{aligned}
\end{equation}
that describes the spin dependent interactions in many magnetic materials satisfactorily where the four most important magnetic interactions in \eqref{Eq:SH} are from left to right, the interatomic exchange \eqref{Eq:SHEX}, magnetic dipole-dipole (\Cref{fig:spin-model}), magnetocrystalline anisotropy and Dzyaloshinskii–Moriya (DM) interactions. 
${\bf e}_K$ is the direction of the anisotropy axis and $K$ and ${\bf D}$ are fitting parameters. 
For a cubic material like bulk Fe, the first two terms in \eqref{Eq:SH} are dominant. 
Unless the shape of the crystal plays a role as it does for thin films, the dipole-dipole interaction can be neglected and we will here concern ourselves mainly with the interatomic exchange interaction.
We will derive two-body and three-body descriptors by analogy with ML-FFs and therefore begin with a brief recapitulation of the GAP-SOAP.  

\begin{figure}[t]
\includegraphics[width=5.6cm]{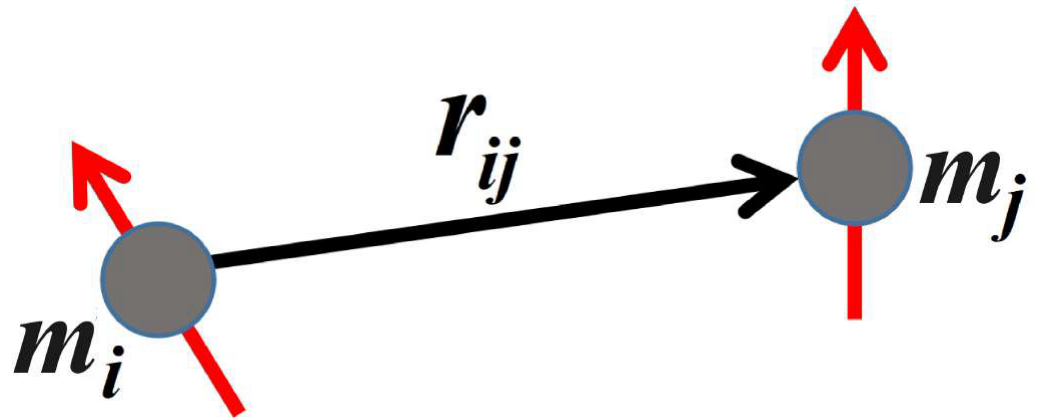}
\caption{\label{fig:spin-model} Two spins ${\bf m}_i$ and ${\bf m}_j$ experience various interactions that depend on their magnitudes $m$ and orientations ${\bf e}={\bf m}/m$ as well as their relative position ${\bf r}_{ij}$. }
\end{figure}

\subsection{Smooth overlap of atomic positions (SOAP)}
\label{ssec:SOAP}

The starting point of recent work on interatomic potentials \cite{Behler:prl07, Bartok:prl10, Bartok:prb13, Jinnouchi:prl19, Jinnouchi:prb19, Drautz:prb19} is the approximation of the potential energy $U$ of a structure with $N_a$ atoms as a sum of local energies $U_i$ 
\begin{equation}
\label{IIA1}
U=\sum_{i=1}^{N_a} U_i.
\end{equation}
where $U_i$ is assumed to be fully determined by the local environment of atom $i$ described by the probability density $\rho_i({\bf r})$ 
\begin{equation}
\label{IIA2}
\rho_i({\bf r}) = \sum_{j=1}^{N_a} f_{\rm cut}({\bf r}_{ij}) g({\bf r}-{\bf r}_{ij}).
\end{equation}
In the SOAP approximation \cite{Bartok:prb13}, the delta functions in \eqref{Eq:rho} are replaced with normalized Gaussians 
\begin{equation}
\label{IIA3}
g({\bf r}- {\bf r}_{ij}) = \frac{1}{(\sqrt{2\pi}\sigma_a)^3}
  \exp \Big(\!\!-\frac{({\bf r}-{\bf r}_{ij})^2}{2\sigma_a^2}\Big)
\end{equation}
and the similarity $S(\rho,\rho^\prime)$ of two structures characterized by probability densities $\rho$ and $\rho'$ is defined as their inner product
\begin{equation}
\label{IIA4}
 S(\rho,\rho') = \int  \rho({\bf r}) \rho'({\bf r}) d{\bf r}.\\    
\end{equation}
Integration of \eqref{IIA4} over all possible rotations $\widehat{R}$ of e.g. $\rho'$
yields the rotationally invariant similarity kernel $k(\rho,\rho^\prime)$
\begin{eqnarray}
\label{IIA5}
\!\!\!\!\!\!\!\!  k(\rho,\rho')  =\int  S({\widehat R}) d{\widehat R} 
&&\equiv \int  S(\rho , {\widehat R} \rho') d{\widehat R} \nonumber \\
&&= \iint  \rho({\bf r}) \rho'({\widehat R}{\bf r}) \, d{\bf r} \, d{\widehat R}                        \end{eqnarray}
By expanding $g({\bf r})$ in plane waves \cite{Kaufmann:jpb89}, $\rho_i$ can be expressed in the spherical harmonics $Y_{lm}(\theta, \phi)$
\begin{eqnarray}
\label{IIA6}
\rho_i({\bf r})=&&\sum_{j=1}^{N_a} f_{\rm cut}({\bf r}_{ij})\frac{1}{(\sqrt{2\pi}\sigma_a)^3}
\exp\Big(\!\!-\frac{ r^2+ r_{ij}^2}{2\sigma_a^2}\Big)  \nonumber\\
&&4\pi \sum_{l=0}^{L_{\rm max}} \sum_{m=-l}^{m=l} \iota_l\Big(\frac{rr_{ij}}{\sigma_a^2}\Big) Y_{lm}^*(\hat {\bf r}_{ij})Y_{lm}(\hat {\bf r}).
\end{eqnarray}
using spherical coordinates ${\bf r}=(r, \theta, \phi)$, the notation ${\bf \hat{r}}= (\theta,\phi)$ for the unit vector and $\iota_l$ is the modified spherical Bessel function of the first kind \cite{Bartok:prb13}. 
On expanding the radial part in a set of orthonormal radial basis functions $\chi_{nl}(r)$  \cite{footnote1}, the atomic density can be rewritten as 
\begin{equation}
\label{IIA7}
\rho_i({\bf r})=\sum_{n=1}^{N_R^l} \sum_{l=1}^{L_{\rm max}} \sum_{m=-l}^{l} c_{nlm}^i \chi_{nl}(r)Y_{lm}(\hat {\bf r}).
\end{equation}

In the GAP framework \cite{Bartok:prl10, Bartok:prb13}, a set of $N$ local reference structures characterized by $\rho_B({\bf r})$ is chosen so that each local energy $U_i$ can be obtained in a kernel expansion \cite{Drautz:prb19} as 

\begin{equation}
\label{IIA8}
U_i=\sum_{B=1}^{N}\alpha_B k({\rho_i},{\rho_B}),
\end{equation}
in terms of the kernels $k(\rho_i,\rho_B)$ between the local structure ${\rho_i}$ of atom $i$ and the ${\rho_B}$ where the fitting parameters $\alpha_B$ are determined by Gaussian process regression \cite{Bartok:prl10, Bartok:prb13}. 
The reference structures ${{\rho_B}}$ are chosen to have minimum overlap with each other by sparsifying the kernel matrix $\{k(\rho_i,\rho_j)\}$ and $N$ is determined iteratively \cite{footnote10}. 

Rotationally invariant two-body and three-body terms, the descriptors $\rho_i^{(2)}(r)$ and $\rho_i^{(3)}(r,s,\theta)$, respectively, can be constructed in terms of $\chi_{nl}$ and $Y_{lm}(\hat {\bf r})$  \cite{Jinnouchi:prb19}.
%
\begin{equation}
\label{IIA9}
\rho_i^{(2)}(r) = \frac{1}{4 \pi} \int \rho_i(r{\bf \hat{r}}) d{\bf \hat{r}}=\frac{1}{\sqrt{4 \pi}} \sum_{n=1}^{} c_n^i \chi_{nl}(r)
\end{equation}
is the two-body {\it radial distribution function} with $c_n^i=c_{n00}^i$ and
\begin{equation}
\label{IIA10}
\begin{aligned}
\begin{split}
&\rho_i^{(3)}(r,s,\theta) = \iint \delta({\bf \hat{r}}.{\bf \hat{s}} -\cos \theta) 
         \rho_i(r{\bf \hat{r}}) \rho_i(s{\bf \hat{s }}) d{\bf \hat{r}} d{\bf \hat{s}} \\ 
&= \sum_{l=1} \sum_{n=1} \sum_{n'=1}\frac{4\pi}{\sqrt{2(2l+1)}} p^i_{nn'l} \chi_{nl}(r)Y_{lm}(\hat {\bf r}) \chi_{nl}(s)Y_{lm}(\hat {\bf s})
\end{split} 
\end{aligned}
\end{equation}
with 
\begin{equation}
\label{IIA11}
p^i_{nn'l}=\sqrt{\frac{8\pi^2}{2l+1}}\sum_{m=-l}^l c^i_{nlm} c^{i*}_{n'lm}.
\end{equation}
is the {\it angular distribution function} \cite{Jinnouchi:prb19} that decomposes the local atomic environment of atom $i$ into triplets involving atoms $i$, $j$ and $k$ and is equivalent to the power spectrum \cite{Bartok:prb13}. 
$\theta$ is the angle subtended by ${\bf r} \equiv {\bf r}_{ij}$ and ${\bf s} \equiv {\bf r}_{ik}$, the position vectors of atoms $j$ and $k$ with respect to atom $i$, respectively. 

The rotationally invariant similarity kernel $k({\rho}, {\rho}')$ becomes 
\begin{equation} 
\label{IIA12}
k({\rho}, {\rho}')=\sum_n c_n c'_n  \,\,\,\,\,\,{\rm and}\,\,\,\,\,\, =\sum_{nn'l} p_{nn'l} p'_{nn'l}
\end{equation}
for two-body and three-body descriptors, respectively where the atom index $i$ is not shown.

\subsection{Smooth overlap of spin orientations (SOSO)}
\label{ssec:SOSO}

By comparison with ML-FFs which address the effective atomic interactions of nonmagnetic materials \cite{Bartok:prb13, Jinnouchi:prb19}, ML-FFs for magnetic systems require an expansion of the phase space to include not only atomic positions but also magnetic moments, a computationally very expensive increase. 
To develop a practical scheme suitable for {\it ab-initio} SD (AISD), we adopt the adiabatic separation of diagonal and off-diagonal components of the density matrix according to which the spin orientation varies slowly in time whereas its magnitude may vary more rapidly \cite{Antropov:prl95, *Antropov:prb96}.
When the rapid, nonadiabatic change of the magnitude $m_i$ of the atomic moment on atom $i$ is neglected, the equation of motion for ${\bf m}_i= m_i{\bf e}_i$ in ASD,
\begin{equation}
\label{IIB1}
\frac{\partial{\bf m}_i}{\partial t} 
= - \gamma {\bf m}_i \times \bigg(-\frac{\partial U}{\partial {\bf m}_i}\bigg) ,
\end{equation}
becomes 
\begin{equation}
\label{IIB2}
\frac{\partial{\bf m}_i}{\partial t} 
 = - \gamma {\bf e}_i \times \bigg(-\frac{\partial U}{\partial {\bf e}_i}\bigg) 
\end{equation}
and the expansion of the phase space can be reduced to a minimum.  
In this adiabatic scheme, $m_i$ can depend implicitly on the orientations of the neighbouring moments $\{ {\bf e}_j \}$ and vary on the same slow time scale. 

To compare the similarity of two spin environments, we will extend the SOAP method, originally designed to distinguish two local structures, to calculate the similarity of two local spin structures in terms of the spin orientations and positions $\{ {\bf e}_i,{\bf r}_i \}$ with what we will call ``Smooth Overlap of Spin Orientations (SOSO)''. 
Because magnetic interactions depend on atomic separations, the SOSO framework will include the smooth overlap of atomic positions as well. 
By analogy with SOAP, we use a Gaussian distribution function to describe the unit moment orientation ${\bf e}_i$ on atom $i$ by 
\begin{equation} 
\label{IIB3}
g({\bf e}-{\bf e}_i) \sim \exp\Big(-\frac{({\bf e}-{\bf e}_i)^2}{2\sigma_s^2}\Big)
=\exp\Big(-\frac{1-{\bf e}\cdot{\bf e}_i}{\sigma_s^2}\Big)
\end{equation}  
where $\sigma_s$ determines the width of the Gaussian distribution of $\bf e$. 
The normalization factor is obtained by integrating the Gaussian function over the surface of the unit sphere with ${\bf e}_i=(0,0,1)$
\begin{eqnarray}
\label{IIB4}
\!\!\!\! \int g({\bf e}-{\bf e}_i) d{\bf e} \sim \!\!\!\!
&&\int\limits_0^\pi \exp \Big(-\frac{1}{\sigma_s^2}\Big) \exp\Big(\frac{\cos \theta}{\sigma_s^2}\Big) d\theta \nonumber\\
= && \exp(-\sigma_s^{-2}) \pi I_0(\sigma_s^{-2}),
\end{eqnarray}
where $I_0(x)$ is the modified Bessel function of the first kind \cite{Abramowitz:65}. 
The normalised Gaussian function 
\begin{equation}
\begin{aligned}
\label{IIB5}
g({\bf e}-{\bf e}_i) 
  =\frac{\exp(\sigma_s^{-2})}{\pi I_0(\sigma_s^{-2})} 
\exp\Big(-\frac{({\bf e}-{\bf e}_i)^2}{2\sigma_s^2}\Big),
\end{aligned}
\end{equation}
represents the spin orientation distribution (SOD) on the unit sphere at site $i$. 
Expanding in spherical harmonics, it can be rewritten as
\begin{eqnarray}
\label{IIB6}
g({\bf e}-{\bf e}_i) = &&\frac{\exp(\sigma_s^{-2})}{\pi I_0(\sigma_s^{-2})} 4\pi \exp(-\sigma_s^{-2}) \nonumber\\
                     && \sum_{l=0}^{L_{\rm max}}  \sum_{m=-l}^l \iota_l(\sigma_s^{-2}) 
                                Y_{lm}^*({\bf e}_i) Y_{lm}({\bf e})  \nonumber\\
=\frac{4}{I_0(\sigma_s^{-2})} && \sum_{l=0}^{L_{\rm max}} \sum_{m=-l}^l \iota_l(\sigma_s^{-2}) Y_{lm}^*({\bf e}_i) Y_{lm}({\bf e}) \nonumber\\
= && \sum_{l=0}^{L_{\rm max}} \sum_{m=-l}^{l} C^i_{lm} Y_{lm}({\bf e}). 
\end{eqnarray}
with 
\begin{equation}
\label{IIB7}
 C^i_{lm} = \frac{4}{I_0(\sigma_s^{-2})} \iota_l(\sigma_s^{-2}) Y_{lm}^*({\bf e}_i) 
\end{equation}

\subsection{SOSO descriptor for interatomic exchange}
\label{ssec:exchange}

We first construct a two-body descriptor for the dominant interatomic exchange interaction in magnetic materials. We do this by generalizing \eqref{IIA2} to include the spin degrees of freedom in terms of the SODs for sites $i$ and $j$
%
\begin{eqnarray}
\label{IIC1}
\rho_i({\bf r},{\bf e}_1,{\bf e}_2)
= \sum_{j=1}^{N_a} f_{\rm cut}( r_{ij}) 
&&g({\bf r}-{\bf r}_{ij}) \nonumber \\
g({\bf e}_1 - {\bf e}_i) 
&&g({\bf e}_2 - {\bf e}_j),
\end{eqnarray}
in which ${\bf e}_1$ and ${\bf e}_2$ are independent unit spin variables.  
Because of the short range of the exchange interaction, we will use the same  cutoff function $f_{\rm cut}( r_{ij})$ as before so only spins within the cutoff distance $r_{\rm cut}$ are considered.

In the classical Heisenberg Hamiltonian \eqref{Eq:SH}, the interatomic exchange energy $E_{\rm ex} = -\frac{1}{2}\sum_{i\neq j}J_{ij}(r_{ij}) \, {\bf m}_i \cdot {\bf m}_j$ where 
$J_{ij}$ only depends on the separation $r_{ij}$ and $E_{\rm ex}$ is invariant under simultaneous rotations of ${\bf m}_i$ and ${\bf m}_j$.
Making use of this symmetry, we obtain the two-body descriptor by integrating the density in \eqref{IIC1} over the unit sphere ${\bf \hat{r}}$ so 
%
\begin{eqnarray}
\label{IIC2}
\rho_i(r,{\bf e}_1,{\bf e}_2)=\int d{\bm{\hat r}} 
\sum_{j=1}^{N_a} f_{\rm cut}( r_{ij}) &&g({ r}-{ r}_{ij}) \nonumber \\
g({\bf e}_1 - {\bf e}_i)
&&g({\bf e}_2 - {\bf e}_j).
\end{eqnarray}
For computational efficiency, the radial part of \eqref{IIC2} is expanded in the normalized spherical Bessel functions $\chi_{nl}(r)=j_l(q_{nl}r)$ and can then be rewritten as
\begin{equation} 
\label{IIC3}
 \rho_i(r)=\frac{1}{\sqrt{4\pi}} \sum_{n=1}^{N_R}c_{n}^i \chi_{nl}(r).  
\end{equation}
The $q_{nl}$ are such that $j_l(q_{nl}r_{\rm cut}) =0$ \cite{Jinnouchi:prb19} and $ c^i_n \equiv c^i_{n00}$. 

To generalize \eqref{IIA5} to include spin, we consider the rotational invariance of ${\bf m}_i . {\bf m}_j$ (${\bf e}_i . {\bf e}_j$). 
The overlap of a local spin environment $\rho_i$ and a rotated environment $\widehat{R}\rho_i'$ defined as
\begin{equation}
\label{IIC4}
 \!\!\!\!\!  S(\widehat{R}) =  \!\! \iiint r^2 dr d{\bf e}_1 d{\bf e}_2 
 \rho_i(r, {\bf e}_1, {\bf e}_2) 
 \rho_i'(r,\widehat{R} {\bf e}_1,\widehat{R} {\bf e}_2) 
 \end{equation}
then becomes \cite{Bartok:prb13} 
\begin{equation}
\label{IIC5}
 \sum_{\substack {n,m_1,m_2 \\
                       m'_1,m'_2 \\
                       l_1,l_2}} 
   \!\!\!\!\!      C^{i^*}_{nl_1l_2m_1m_2} {C'}^i_{nl_1l_2m'_1m'_2} 
         D_{m_1m'_1}^{l_1} (\widehat{R})D_{m_2m'_2}^{l_2}(\widehat{R})   
\end{equation}
where $C^i_{nl_1l_2m_1m_2}=\frac{1}{\sqrt{4\pi}}c^i_{n}C^i_{l_1m_1}C^i_{l_2m_2}$ in terms of $C^i_{lm}$ \eqref {IIB6} and $c^i_{n}$ (\ref{IIA9},\ref{IIC3}) and we use the relation
\begin{equation}
\label{IIC6}
 Y_{lm'}(\widehat{R} {\bf \hat e})=\sum_{m=-l}^l Y_{lm}({\bf \hat e}) D_{m'm}^{l}(\widehat{R})
\end{equation}
where $D_{m'm}^{l}(\widehat{R})$ is the unitary Wigner rotation matrix for the rotation $\widehat{R}$ that satisfies the relation ${{\bf D}^l}^{\dagger}{\bf D}^{l}={\bf I}$ \cite{Wigner:59}. 

\begin{widetext} 
The similarity kernel including spin can be derived as
\begin{eqnarray}
\label{IIC7}
k(\rho_i,\rho'_i)=\int S(\widehat{R})dR  
=\sum_{\substack {n,m_1,m_2 \\
                 m'_1,m'_2 \\
                 l_1,l_2}} C^{i^*}_{nl_1l_2m_1m_2} {C'}^{\,i}_{nl_1l_2m'_1m'_2} 
                 \int  D_{m_1m'_1}^{l_1} (\widehat{R})D_{m_2m'_2}^{l_2}(\widehat{R}) d\widehat{R} \nonumber \\
= \sum_{\substack {n,m_1,m_2 \\
                   m'_1,m'_2 \\
                   l_1,l_2}} C^{i^*}_{nl_1l_2m_1m_2} {C'}^{\,i}_{nl_1l_2m'_1m'_2} 
  (-1)^{m_1-m'_1} \int D_{-m_1,-m'_1}^{{l_1}^{*}} (\widehat{R}) 
    D_{m_2m'_2}^{l_2}(\widehat{R}) d\widehat{R} \nonumber \\ 
= \sum_{\substack {n,l,m_1,m'_1}} \frac{8\pi^2(-1)^{m_1+m'_1}}{2l+1}  C^{i^*}_{nllm_1,-m_1} {C'}^{\,i}_{nllm'_1,-m'_1}    
\end{eqnarray}
\end{widetext}
using the property $D_{m m'}^l (\widehat{R}) = (-1)^{m-m'} D_{-m, -m'}^{l^*} (\widehat{R})$,
$d\widehat{R}$ indicates integration over all possible rotations \cite{Bartok:prb13} and 
\begin{equation}
\label{IIC8}
Y^*_{lm}(r)=(-1)^m Y_{l,-m}(r).
\end{equation}
We have now given the full expression for the kernel of the two-body term of the exchange interaction. 
Since the latter does not depend on the direction ${\bf r}_{ij}$, only a radial dependence on  $r_{ij}$ is left. 

Summing the coefficients in \eqref{IIC7} over $m_1$ allows us to rewrite the similarity kernel in the same form as \eqref{IIA12}
\begin{equation}
\label{IIC9}
k(\rho_i,\rho'_i)= \sum_{n,l}  \widetilde{C}_{nl}^{i^*} \widetilde{C}'^{\,i}_{nl}.
\end{equation} 
in terms of the coefficients
\begin{widetext}
\begin{subequations}
\label{IIC10}
\begin{align}
\widetilde{C}^i_{nl}= \sum_{m_1} (-1)^{m_1}\sqrt{\frac{8\pi^2}{2l+1}}   C^i_{nllm_1,-m_1} =\sum_{m_1} (-1)^{m_1} \sqrt{\frac{8\pi^2}{2l+1}} \frac{1}{\sqrt{4\pi}}  c^{i}_{n00} C^i_{lm_1}C^i_{l,-m_1}                          \label{IIC10a} \\
=\sum_j^{N_a}\sqrt{\frac{2\pi}{2l+1}}c_{n00}^i 
\Big(\frac{4}{I_0(\sigma_s^{-2})} \Big)^2  
\sum_{m_1=-l}^l \iota_l^2 (\sigma_s^{-2}) Y_{lm_1}^*({\bf e}_i) Y_{lm_1}({\bf e}_j) \label{IIC10b} \\  
=\sum_j^{N_a}\sqrt{\frac{8\pi}{2l+1}}c_{n00}^i    \Big(\frac{4}{I_0(\sigma_s^{-2})}\Big)^2  
 \iota_l^2(\sigma_s^{-2}) P_l(\cos \theta_{ij}), \label{IIC10c} 
\end{align}
\end{subequations}
where $P_l(\cos \theta)$ is a Legendre polynomial, $\theta_{ij}$ denotes the angle between spins $i$ and $j$ and the prefactor $\frac{8\pi^2}{2l+1}$ in \eqref{IIC7} is decomposed into $\sqrt{\frac{8\pi^2}{2l+1}}\times\sqrt{\frac{8\pi^2}{2l+1}}$ and absorbed into $\widetilde{C}^i_{nl}$ and $\widetilde{C}'^{\,i}_{nl}$. 
\end{widetext}

It should be noted that the kernel does not integrate the square of $S(\widehat{R})$ as in the case of GAP-SOAP \cite{Bartok:prb13}. 
Motivated by the success of the Heisenberg Hamiltonian \eqref{Eq:SH} in describing interatomic exchange, we assume that the main contribution to the exchange interaction has two-body form and this descriptor is implemented for testing in \Cref{Sec:train-test}. 
Should it be necessary to increase the accuracy, the descriptor can be extended to higher order terms. 
For three-body or higher order terms, the corresponding kernel cannot be constructed by integrating $ |S(\widehat{R})|^n $ over all possible rotations as in GAP-SOAP, where $n=2$ refers to the three-body term descriptor. 
A three-body descriptor is derived for the interatomic exchange interaction in \Cref{Appendix:A}; the magnetic dipole-dipole interaction is considered in \Cref{Appendix:B}. 

\section{Training and testing}
\label{Sec:train-test}

Ultimately, we want to use DFT energies to train the ML system. 
Because these total energies are very large, it is important to choose suitable reference states to be able to focus on the relatively small changes in energy $\Delta E$  \cite{footnote5} when the spin orientations vary.
For this, we choose the collinear spin configuration with the same spatial coordinates $\{{\bf r}_i\}$ 
\begin{equation} 
\label{III1}
\Delta E(\{{\bf m}_i,{\bf r}_i\})   
  =  E(\{{\bf m}_i,{\bf r}_i\})  - E_{\rm FM}(\{{\bf r}_i\}), 
\end{equation} 
where $E$ and $E_{\rm FM}$ are total energies for noncollinear and collinear ferromagnetic systems, respectively and the GAP-SOAP ML-FF framework can describe $E_{\rm FM}(\{{\bf r}_i\})$ very well \cite{Jana:prb23}. 
To test the performance of the SOSO descriptor, $\Delta E$ will be used to train the ML model using kernel regression as in the GAP framework with the two-body descriptor for interatomic exchange from \Cref{ssec:exchange}
with a view to decomposing the total energy into local energies $\Delta E_i$
%
\begin{equation}
\label{III2}
\Delta E(\{{\bf m}_i,{\bf r}_i\}) =\sum_{i=1}^{N_a} \Delta E_i
\end{equation} 
by analogy with \eqref{IIA1}. 

To fit the potential energy surface $\Delta E(\{{\bf m}_i,{\bf r}_i\})$ using the GAP framework \cite{Bartok:prl10, Bartok:prb13}, we begin by describing the local spin structures $\rho_i$ in the training set with the two-body descriptor for interatomic exchange from \Cref{ssec:exchange}. 
A set of local reference configurations $\{\rho_{B}\}$ whose spin structures have minimum overlap with each other is then selected by sparsifying the kernel covariance matrix $\{k(\rho_i,\rho_j)\}$. 
Finally, $\Delta E(\{{\bf m}_i,{\bf r}_i\})$ is obtained in the form \eqref{III2} using the training sets to fit the ML model with kernel regression yielding a set of coefficients $\alpha_B$ as in \eqref{IIA8} making predictions of the total energy for new spin structures possible.

Once the ML model for $\Delta E(\{{\bf m}_i,{\bf r}_i\})$ is trained, the exchange field ${\bf h}_i$ can be obtained analytically by taking the derivative of the total energy with respect to ${\bf m}_i$. 
Because the magnitudes $\{ m_i \}$ are not included in the present model, only the perpendicular part of the exchange field ${\bf h}_i$ on each site can be obtained analytically from
%
\begin{eqnarray}
\label{III3}
{\bf h}_i^\perp 
  &&= -\frac{1}{m_i}\frac{\partial{\Delta E(\{{\bf m}_i,{\bf r}_i\})}}{\partial{{\bf e}_i}}  \nonumber\\
  &&= -\frac{1}{m_i}\frac{\partial{ \sum_{B=1}^{N}\alpha_B k({\rho_i},{\rho_B})}}{\partial{{\bf e}_i}}.
\end{eqnarray} 
because the kernel $k({\rho_i},{\rho_B})$ is a functional of the spin orientation ${\bf e}_i$ as seen in \eqref{IIC9} and \eqref{IIC10}. 

Though $m_i$ is known in calculations using Heisenberg Hamiltonians or DFT, it does not enter the SOSO model so is not known when we want to extract ${\bf h}_i^\perp$ for some arbitrary atom-spin configuration; instead a fixed magnitude of the local magnetic moment of spin $i$ ($m_i \equiv|{\bf m}_i|$) will be used to evaluate \eqref{III3}. 
While ${\bf h}_i^\perp$ induces torque on the spin, the longitudinal part of the effective field which governs its magnitude is not accounted for in SOSO. 
In the ``adiabatic spin approximation'' (ASD) \cite{Antropov:prb96}, the size of the magnetic moment $m_i$ is the fast degree of freedom relative to ${\bf e}_i$.  
In principle, $m_i$ could also be learned with the GAP-SOSO model, in which $m_i$ can be uniquely determined by the local spin environment of spin $i$. However, this is beyond the scope of the present publication. 
 
To train the GAP-SOSO model constructed in the previous section, there are three steps: (i) generating noncollinear spin structures and dividing them into training and test sets; (ii)  calculating the ``exact'' total energies and exchange fields for both sets; (iii) training the GAP-SOSO model with the training sets and making predictions for the test structures. The performance of the ML model will be evaluated by comparing the total energies and exchange fields ``predicted'' by the model for these structures with the results of the ``exact'' calculations.

\subsection{Tests with a Heisenberg Hamiltonian}
\label{ssec:Test-Heisenberg}

Our first test of the model is for noncollinear spin structures described by the sets of random unit vectors $\{{\bf e}_i\}$ whereby spins $m{\bf e}_i$ on a bcc lattice with the experimental lattice constant $a=2.8665\,$\AA{} \cite{WangH:prb10} and the GGA spin magnetic moment $m=2.23 \, \mu_{\rm B}$ interact via the exchange interaction $J_{ij}(R)>0$ which only depends on the distance $R$ between pairs of spins
\begin{equation}
\label{IIIA1}
\Delta E = -\frac{1}{2}\sum_{i\neq j}J_{ij}(R) \, {\bf m}_i \cdot {\bf m}_j - E_{\rm FM}.
\end{equation}
For a fixed bcc lattice, the separations of first, second and third nearest neighbours are $ \sqrt{3}/2=0.866$, $1$ and $\sqrt{2}=1.414$, respectively, in units of the lattice constant $a$.  
Values of $J(R)$ in the phenomenological Hamiltonian \eqref{IIIA1} are taken from Table~I of Ref.~\cite{WangH:prb10} for the first three shells of neighbours. 
To describe the exchange interaction when atoms are displaced from their lattice sites, we interpolate $J(R)$ using the RKKY-like form 
\begin{equation}
\label{IIIA2}
J(r)=\left\{
     \begin{array}{lr}
       \frac{c}{r^3}  \sin{(k r+\phi)}, & {\rm if} \quad r \leq r_c \vspace{1ex} \\ 
         0, & {\rm otherwise}. 
     \end{array}     
      \right.
\end{equation}
in which $r$ denotes the distance between Fe atoms and $r_c$ represents a cutoff length for the exchange interaction.
$r_c$ is set to be 1.45 times the lattice constant. 
Fitting yields values of the amplitude $c=0.35\,$eV/\AA$^3$, of the wave vector $k=1.55\,$\AA$^{-1}$, and of the phase factor $\phi=-0.97\pi$.  
Although the interpolated values of $J(R)$ decay only slowly and can exceed the value for third nearest-neighbours \cite{WangH:prb10}, we will only be interested in using $J(r)$ for model calculations where interactions between nearest-neighbour (nn) and next-nearest-neighbour (nnn) atoms are dominant. 

Later on we will replace the phenomenological Hamiltonian \eqref{IIIA1} with parameter-free DFT calculations.
The latter are computationally very expensive for noncollinear metallic magnetic systems because a very dense sampling of reciprocal space and many iterations are required to calculate magnetic moments and total energies with sufficient accuracy. 
This means that the DFT calculations are limited to quite small training sets of supercells containing only $16 = 2(2 \times 2 \times 2$) or $54=2(3 \times 3 \times 3$) atoms.

\subsubsection{Fixed $m_i$ and $r_i$}
\label{sssec:fixed}

To gain experience with such necessarily small systems, we generated 2600 noncollinear spin structures of bcc Fe in 16 atom supercells (a 2$\times 2 \times$2 simple cubic supercell with a basis of two Fe atoms), with $\{{\bf e}_i\} \equiv \{\theta_i,\phi_i\}$ expressed in spherical coordinates distributed uniformly in the range $0<\theta<\pi$ and $0<\phi<2\pi$ and chosen randomly for training and testing.
100 configurations were used to train the model while the remaining 2500 were used for testing.

The total energy of each spin configuration was calculated using \eqref{IIIA1} for which the effective exchange field is 
\begin{equation}
\label{IIIA3}
{\bf h}_i = \frac{1}{2}\sum_{j \neq i}J_{ij}(R) \, {\bf m}_j. 
\end{equation}
In the ML model, a cut-off radius of $r_{\rm cut}=7\,$\AA\ was used to calculate the exchange interaction between Fe atoms. 
For the spherical harmonics in \eqref{IIC10}, maximum values of $n_{\rm max}=12$ and $l_{\rm max}=6$ were chosen and our results are insensitive to further increases in the $r_{\rm cut}$, $n_{\rm max}$, and $l_{\rm max}$ ``hyperparameters''. 
As shown in \Cref{fig:Heisenberg}, the SOSO model ``predicts'' the total energy and the transverse exchange field ${\bf h}_i^{\perp}$ essentially perfectly with just 100 training configurations. 
The root-mean-square-error (RMSE) on the predictions for the 2500 random spin configurations shown in \Cref{fig:Heisenberg}(a) is only 0.08  meV/atom. 
For the transverse exchange field on each atom, the RMSE of the magnitude shown in \Cref{fig:Heisenberg}(b) is 0.13\,meV/$\mu_B$ while the distribution of the deviations of the predicted and ``exact'' orientations shown in the inset are seen to be very small, less than $\sim 0.2^{\circ}$ . 
The test sets in the Heisenberg model consist of completely random spin configurations with average angles between nearest neighbour spins of order $\pi/2$, corresponding to an energy of around 7~eV for the 16 atom Fe supercell taking the FM state as reference. 

\begin{figure}[t]
\includegraphics[width=8.6cm]{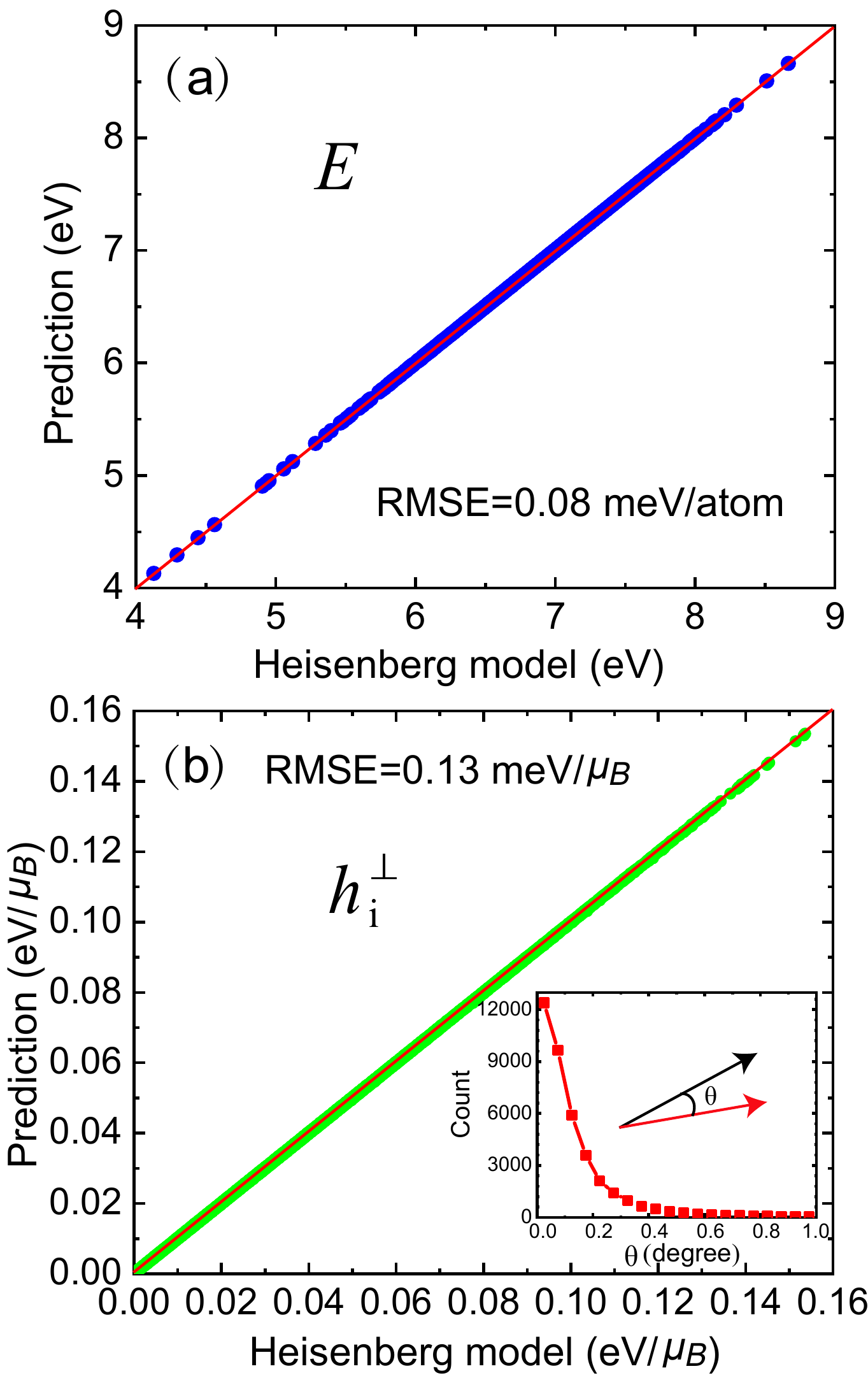}
\caption{\label{fig:Heisenberg}(a) Total energy and (b) transverse exchange field predicted by the ML-EF model compared with the ``exact'' values  given by the Heisenberg Hamiltonian \eqref{IIIA1} with fixed magnetic moment in a 2$\times$2$\times$2 simple cubic supercell of bcc Fe. 
The inset in (b) shows the distribution of angles between the predicted transverse exchange field (red arrow) and the exact ones (black arrow) for 2500 noncollinear spin configurations.} 
\end{figure}

\begin{figure}[t]
\includegraphics[width=8.6cm]{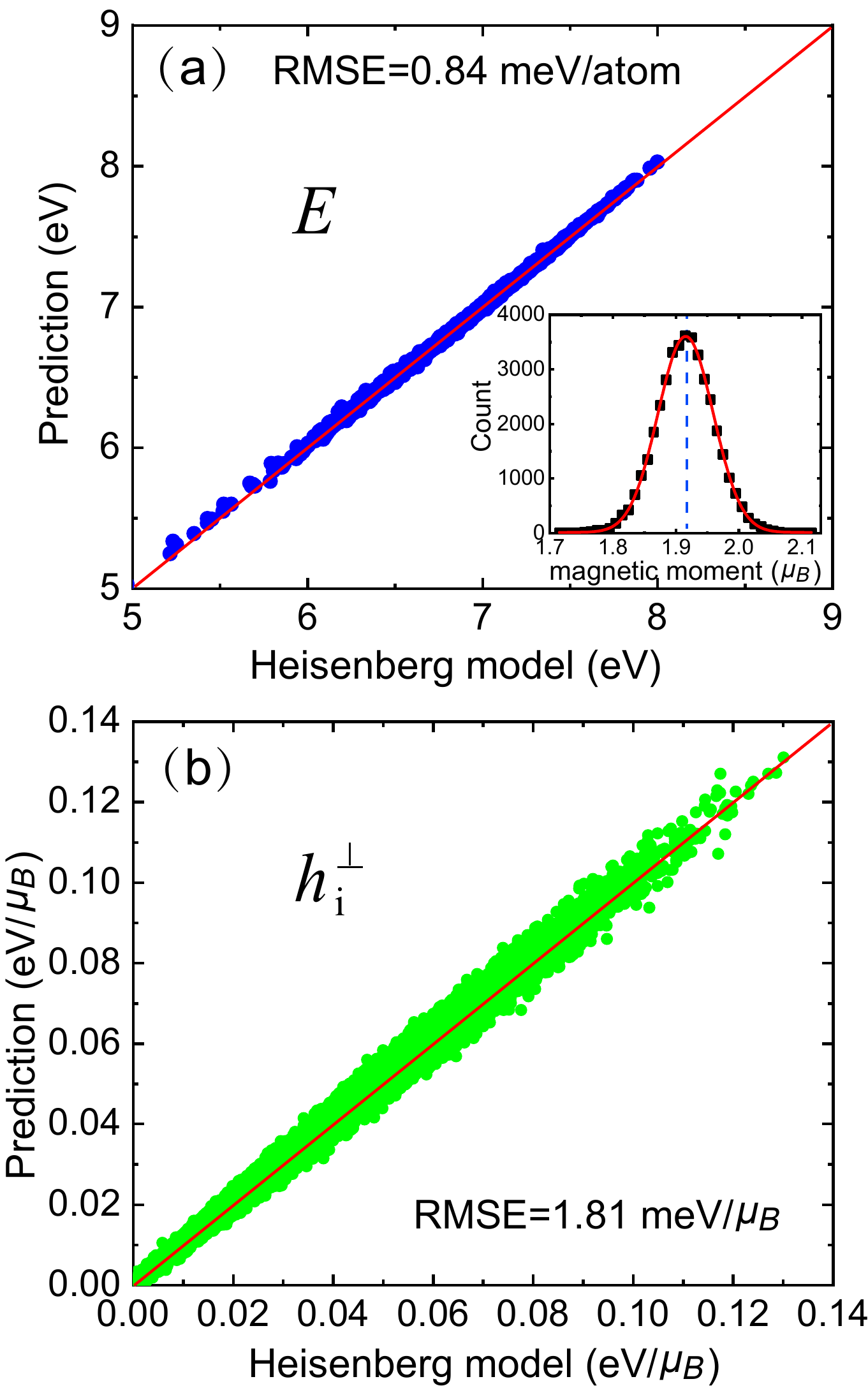}
\caption{\label{fig:HHm} (a) Total energy and (b) transverse exchange field predicted by the ML-EF model compared with the ``exact'' values given by the Heisenberg Hamiltonian \eqref{IIIA1} when the magnetic moment on site $i$ depends on the orientation of the magnetic moments of its nearest neighbours as described by \eqref{IIIA3}.
The distribution of magnetic moments in the 2500 randomly generated noncollinear configurations is shown in the inset in (a) where the red curve is the Gaussian fit to the data with a mean value of $\overline{m}=1.915 \, \mu_B$ and standard deviation $\sigma_m=0.044\,\mu_B$.}
\end{figure}

The SOSO descriptor is essentially exact for the model Hamiltonian \eqref{IIIA1}. 
Before implementing it for a DFT energy functional, we want to address two important features of \Cref{fig:GGA_EMa}(b). 
First, in both the FM and AFM states the magnetic moment depends on the lattice constant $a$ which itself depends on temperature. 
Second, the magnetic moment $m$ of bcc Fe at $T=0\,$K depends on whether all the spins are parallel in the lowest energy FM state or antiparallel as in the AFM state in which the magnetic moment of the Fe atom at the centre of the simple cubic unit cell is equal and opposite to that of the Fe atom at the origin.
At the room temperature experimental volume corresponding to $a=2.8665 \,$\AA\ \cite{Basinski:prsla55} the local GGA spin magnetic moment per atom is 2.23$\, \mu_B$ for the FM and only 1.60$\, \mu_B$ for the AFM states. 
We want to examine to what extent local magnetic moments are described implicitly by the local spin environment $\{ {\bf e}_i, {\bf r}_i\}$ in terms of how well we can predict the total energy without explicitly knowing  $\{ m_i\}$.

\subsubsection{Variable $m_i$}
\label{sssec:TH-m}

To see how well the SOSO model works when the size of atomic moments depends on the local environment, we modified the Heisenberg Hamiltonian so that the moment of spin $i$ depends linearly on the angles it makes with its nearest neighbours (nn) as  
\begin{equation}
\label{IIIA4}
\begin{aligned}
 m_i= 2.23 - \sum_j^{\rm nn} \frac{\cos^{-1}({\bf e}_i \cdot {\bf e}_j)}{n\pi} \times 0.63 ,
\end{aligned}
\end{equation}
in which $n$ is the number of nearest neighbours of spin $i$ (8 for bcc) and the constant 0.63 (=2.23-1.60) comes from the GGA moment variation between FM and AFM states with an angle of $\pi$. 
\Cref{fig:HHm}(a) shows that the spin-configuration-dependent magnetic moment increases the RMSE of the predicted energy to 0.84 meV/atom, small compared to the total energy but no longer essentially zero. 
The RMSE of the transverse exchange field, \Cref{fig:HHm}(b), increases from 0.13 to 1.81~meV/$\mu_B$, which we attribute to the use of a fixed magnetic moment to predict ${\bf h}_i^\perp$ in \eqref{III3}. 
Although the variation in $m_i$ in \eqref{IIIA4} is almost 25\%, this variation is not reflected in $h_i^\perp $. 
The distribution of magnetic moments that results from randomly generating 2500 spin configurations is seen in \Cref{fig:HHm}(a) (inset) to be Gaussian with a mean value of $\overline{m}=1.915 \mu_B$ and standard deviation $\sigma_m=0.044\mu_B$, which is around $2.2\%$ of $m_i$.
This is consistent with our expectation that the variance of $h_i^\perp$ should be proportional to that of $m_i$ and accounts for the observed RMSE very well. 
The total energy of the Heisenberg model is thus predicted remarkably well even though the moment magnitudes are not included (explicitly) in the SOSO model. 

\begin{figure}[t]
\includegraphics[width=8.4cm]{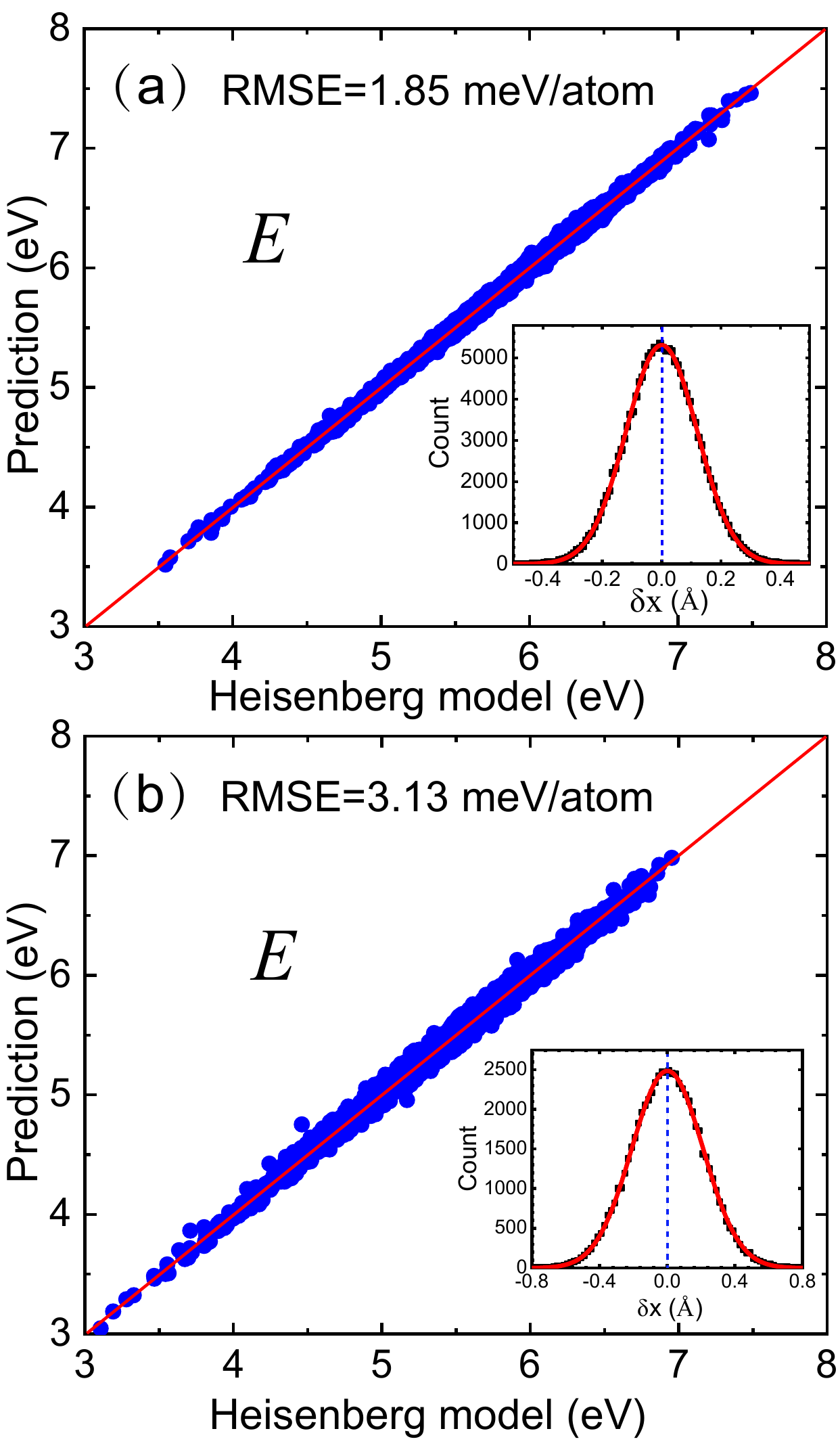}
\caption{\label{fig:spin-lattice} Comparison of the total energies predicted by the ML-EF model with those determined explicitly with the Heisenberg Hamiltonian with atomic displacements in a 16 atom supercell of bcc Fe. 
The distribution of atomic displacements is assumed to be Gaussian with (a) $\sigma_r = 0.11$ \AA{} and (b) $0.21$ \AA{} in \eqref{IIIA6}. }
\end{figure}

We can obtain a more accurate estimate of $h_i^\perp$ by using the exact $m_i$ from \eqref{IIIA4}. 
On doing so we find a reduction of the RMSE from 1.81 to 1.25 meV$/\mu_B$.
We conclude that the latter value must result from the accuracy of calculating $\partial \Delta E / \partial{\bf e}_i$ using the SOSO model.
It can be improved somewhat by using larger training sets; the RMSE can be roughly reduced by 50\% with training sets ten times larger, 1000 rather than 100.
As pointed out in \Cref{ssec:SOSO}, the factors $m_i$ cancel in the equation of motion for the spin so that in the ASD, the spin dynamics are given correctly by solving 
\begin{equation}
\label{IIIA5}
\frac{\partial{\bf m}_i}{\partial t} 
= \gamma {\bf e}_i \times \frac{\partial \Delta E}{\partial {\bf e}_i} 
\end{equation}
where the energy gradient can be determined from the magnetic PES in $\{ {\bf e}_i \}$ space.

\subsubsection{Variable atomic coordinates ${\bf r}_i$}
\label{sssec:TH-r}

The noncollinear spin systems considered so far are for atoms (and spins) on fixed lattice sites. 
To evaluate the model's performance for systems with thermally disordered spin orientations and positions, we introduce a three-dimensional Gaussian distribution 
\begin{equation}
\label{IIIA6}
    P(\Delta_r)= \frac{1}{  (2\pi)^{3/2} \sigma_r^3 } 
      \exp\Big(-\frac{\Delta_r^2}{2\sigma_r^2}\Big)
\end{equation}
of displacements $\Delta_r$ of spins from their equilibrium lattice sites for the noncollinear spin system where $\sigma_r$ is the width of the distribution. 
We choose $\sigma_r$ in \eqref{IIIA6} to be 0.11 \AA\ and 0.21 \AA\ corresponding to temperatures of 300K and 1000K, respectively, according to \cite{Ziman:72}
\begin{equation}
\label{IIIA7}
\sigma_r^2 = \frac{9 \hbar^2 T}{Mk\Theta^2} 
\end{equation}
where the Debye temperature $\Theta=420\,$K for Fe \cite{A&MSSP}. 
Although the expression \eqref{IIIA7} is formally only valid above the Debye temperature, in practice it holds for much lower temperatures \cite{footnote8}.
Keeping the magnitude of the magnetic moment fixed at 2.23 $\mu_B$, random spin configurations were generated with $0<\theta<\pi$ and $0<\phi<\pi$.
The variation of both spin orientation and position leads to a large phase space of spin configurations. 
We use 1000 samples as training sets and keep 2500 samples as test sets. 

\Cref{fig:spin-lattice} shows that the ML-EF model describes the total energy of a system with a distribution of spin orientations and atomic displacements quite well.
The RMSE increases from 1.85 to 3.13 meV/atom when the RMS displacement $\sigma_r$ is increased from 0.11 to $0.21\,$\AA. 

\subsection{Tests for constrained DFT calculations}
\label{ssec:Test-DFT}

In reality, the exchange interaction will not only depend on the separations of spins but also on the local noncollinear spin configurations in a more complicated way than suggested by the simple linear dependence assumed in \eqref{IIIA4}. 
Assuming that DFT-GGA describes spin-spin interactions accurately, we proceed to test the SOSO descriptor using DFT-GGA training and testing data. 

To generate representative noncollinear spin configurations at temperature $T$, we used the Uppsala ASD (UppASD) package \cite{Skubic:jpcm08, Eriksson:17} to perform SD simulations with a time step of $0.1 \,$fs in 16-atom supercells of bcc Fe by numerically solving the atomistic Landau-Lifshitz-Gilbert equation. 
We use a Langevin thermostat to maintain the desired temperature (300K and 1000K) and the same exchange parameters as in the previous section to generate effective exchange fields \eqref{IIIA3}. 
Once equilibrium was reached at the chosen temperature, 10000 spin structures were selected at intervals of $10\,$fs. 
Because of the large computational cost of constrained DFT calculations for noncollinear spin configurations, only 50 spin structures were picked randomly from these 10000 spin structures for each of two temperatures, 300 K and 1000 K, of which half were used for training while the other half were reserved for testing. 

The $\alpha$th spin configuration obtained from the UppASD calculations is characterized by the set of orientations $\{{\bf e}_i^\alpha \}$ where $i$ (1-16) is the atom index in the $2 \times 2 \times 2$ simple cubic supercell of bcc Fe. 
Because these configurations are in general not local minima of the spin density functional, the Kohn-Sham equations cannot be simply solved self-consistently for a configuration $\{{\bf e}_i^\alpha \}$; each spin orientation must be constrained so it is not changed by iteration. 
This is done using a procedure implemented in {\sc vasp} (following a methodology developed by Ma and Dudarev \cite{Ma:prb15}) whereby an energy penalty $E_p$ is added to the total energy $E_0$ when a magnetization ${\bf m}({\bf r})$ deviates from the required orientations $ \{{\bf e}_i \}$ \cite{footnote7} 
\begin{equation}
\label{IIIB1}
E= E_0 + E_p = E_0 + \sum_i \lambda_i \big[ {\bf m}_i - {\bf e}_i ({\bf e}_i. {\bf m}_i) \big]^2 
\end{equation}
where the atomic moment ${\bf m}_i$ is defined as 
\begin{equation}
\label{IIIB2}
{\bf m}_i =\int_{\Omega_i} {\bf m}({\bf r}) F_i(|{\bf r}-{\bf r}_i|) d{\bf r},
\end{equation}
$\Omega_i$ is a sphere centred on atom $i$ and $F_i$ projects the magnetization ${\bf m}({\bf r})$ onto this atom \cite{Ma:prb15}.
The penalty introduces an additional potential on each atom $i$   
\begin{equation}
\label{IIIB3}
V_i({\bf r}) = 2 \lambda_i \big[  {\bf m}_i - {\bf e}_i ({\bf e}_i. {\bf m}_i) \big]
         \cdot {\bm \sigma} F_i(r) \\
         = - {\bf b}_p ({\bf r}) \cdot {\bm \sigma}
\end{equation}
and 
\begin{equation}
\label{IIIB4}
{\bf h}^{\rm Con}_i =\frac{1}{2} \int_{\Omega_i} {\bf b}_p ({\bf r})   d{\bf r}
\end{equation}
where $\lambda_i \rightarrow \infty$ represents the parameter used in the constrained DFT calculations. 
${\bf h}^{\rm Con}_i$ should cancel the perpendicular part of the exchange field on spin ${\bf m}_i$ to yield a local minimum of the total energy. Therefore, the perpendicular part of the exchange field ${\bf h}^\perp_i$ \eqref{III3} about which the $i$th spin precesses can be obtained as the negative of the constraining field ${\bf h}^{\rm Con}_i $.

\begin{figure*}[t]
\includegraphics[width=17cm]{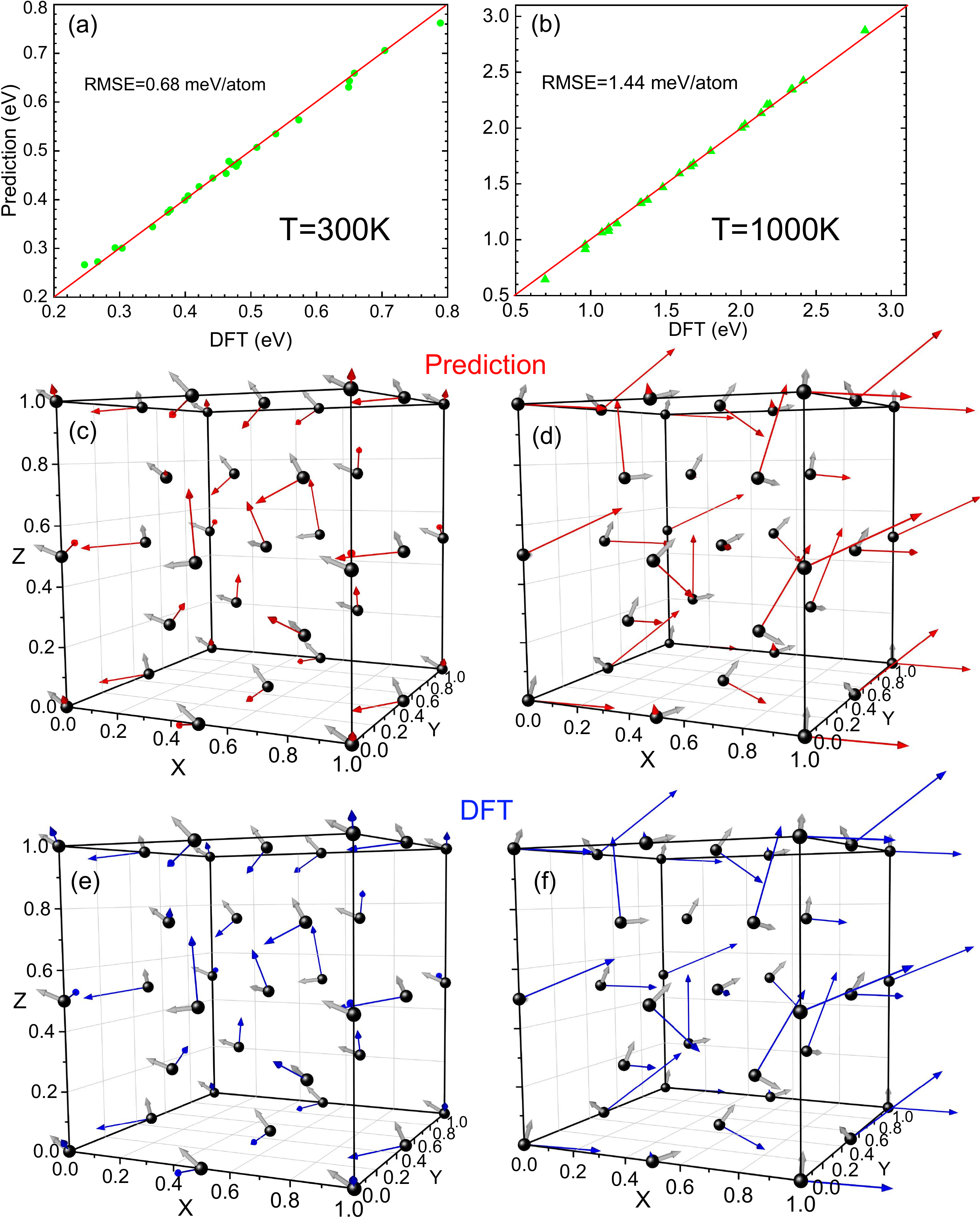}
\caption{\label{fig:300-1000} (Top row) Comparison of total DFT energies calculated for 25 noncollinear test spin configurations described in 16 atom supercells of bcc Fe with those predicted using the ML model trained with a separate set of 25 non-collinear spin configurations at (a) 300K and (b) 1000K, respectively. 
All energies are given with respect to the energy of the corresponding collinear system and RMSE is the root mean square error.  
Comparison of the perpendicular component of the exchange fields ${\bf h}_i^\perp$ (red arrows) on the Fe atoms predicted by the ML-EF model (middle row) for a single spin configuration $\{ {\bf e}_i\}$ of the test set (indicated with grey arrows) with (bottom row) the same quantity (blue arrows) calculated ``exactly'' with constrained DFT calculations at 300K (left column) and at 1000K (right column), respectively. 
The unit of the exchange field is eV/$\mu_B$ and is scaled by a factor of 4 for visualization purposes. The black balls indicate the Fe atoms.}
\end{figure*}

For each set $\{{\bf e}_i^\alpha \}$, we used the {\sc vasp} code to converge the total energy of the noncollinear spin system to $10^{-7}$eV/atom using a plane-wave basis with a cutoﬀ energy of 500 eV and a 6$\times 6 \times$6 $\Gamma$-centered reciprocal space mesh requiring of order 1000 (respectively 4000) core hours of computation for a 2$\times 2 \times$2 (respectively 3$\times 3 \times$3) supercell.  
To constrain the spin direction, $\lambda$ is set to be 30 with the energy penalty converging to less than $0.5 \times 10^{-4}$eV/atom. 
With these computational parameters, the atomic moments in the constrained noncollinear calculations are converged to better than $\sim$1\%. 
The DFT energies of the noncollinear spin systems used for training and testing were defined with respect to the corresponding collinear ferromagnetic system as defined in \eqref{III1}. Because of the small size of the magnetocrystalline anisotropy energy (MAE) for cubic systems, \Cref{TableI}, spin orbit coupling was not included.

\subsubsection*{Energies}

\begin{figure}[b]
\includegraphics[width=8.6cm]{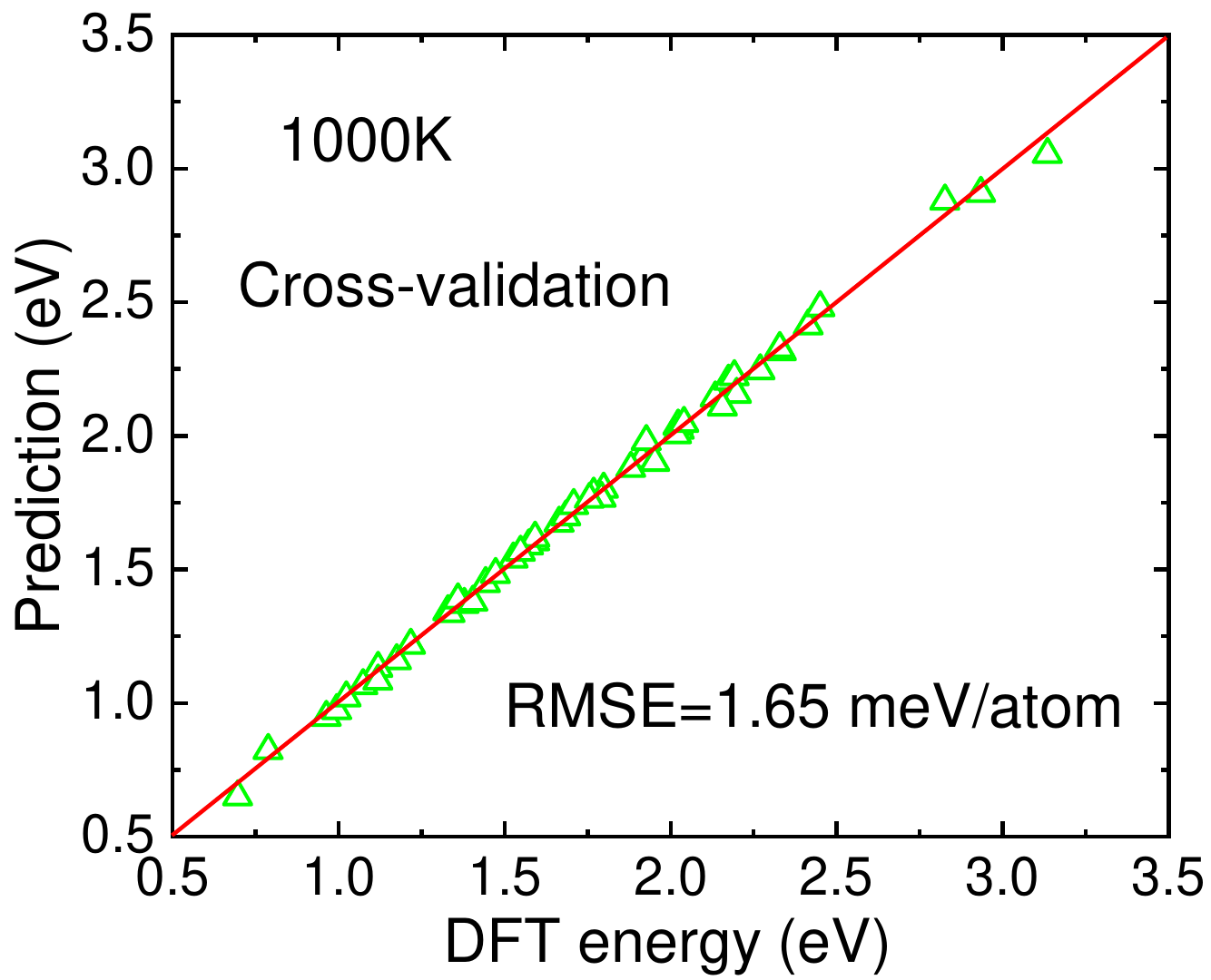}
\caption{\label{fig:CV}Comparison of total DFT energies calculated for the 50 noncollinear configurations of 16-atom supercells of Fe with those predicted using the ML model at 1000K where the 50 configurations were partitioned into five groups of ten rather than two groups of 25. 
For each test group of ten the remaining 40 configurations were used for training.
}
\end{figure}

Choosing a temperature, we train the SOSO model using 25 training configurations and use it to predict the energies of the remaining 25 test configurations. 
The average energy of the 25 test configurations increases from $\sim 0.5\,$eV at 300~K in \Cref{fig:300-1000}(a) to $\sim 1.7\,$eV at 1000~K in \Cref{fig:300-1000}(b) where we expect the average energy of classical spins with three rotational degrees of freedom to be $\frac{3}{2}k_BT$ and the total thermal energy to be $\sim 16 \times \frac{3}{2} k_BT$, which is 0.6 eV for 300K and 2eV for 1000K. 
The large spread in energies reflects the smallness of the system under study.
The energies predicted by the model are compared to the exact DFT energies in the top row of \Cref{fig:300-1000}. 
The results demonstrate that the ML-EF PES trained with a very small set of 25 spin structures predicts the total energies of noncollinear spin structures remarkably accurately. 
The RMSE of the total energy is less than 1 meV/spin at 300 K, increasing to 1.44 meV/spin at 1000K. 
The larger RMSE at 1000 K suggests the need for a more extensive training set to represent the phase space of spin structures at higher temperatures more faithfully. 

To obtain the complete magnetic PES of bcc Fe may require training the model with sets that sample a larger part of the entire phase space ranging from 0K to the phase transition temperature. 
The energy scale of the DFT calculations ($\sim$ 0.5 eV for 300K to $\sim$2 eV for 1000K) are much smaller than in the Heisenberg model (4-10 eV in \Cref{fig:Heisenberg}(a)), primarily because the angles between spins in the Heisenberg model were randomly distributed in the range [0,$\pi$] whereas the spin configurations obtained from the UppASD simulations correspond to finite temperatures and the average angles range from $0.14 \pi$ for 300~K to $ 0.33 \pi$ for 1000 K.

The partitioning of the 50 selected spin structures into equal sized training and testing sets is arbitrary.
For the 1000K data, we performed a 5-fold cross-validation by partitioning the fifty samples into five groups of ten. 
Taking each group in turn as the test set, the remaining four groups were used for training yielding the fifty data points shown in \Cref{fig:CV}. 
Using all 50 predictions and true values resulted in a RMSE energy of 1.65~meV/atom compared to 1.44~meV/atom in \Cref{fig:300-1000}(b). 

\subsubsection*{Moments}

\begin{figure}[b]
\includegraphics[width=8.6cm]{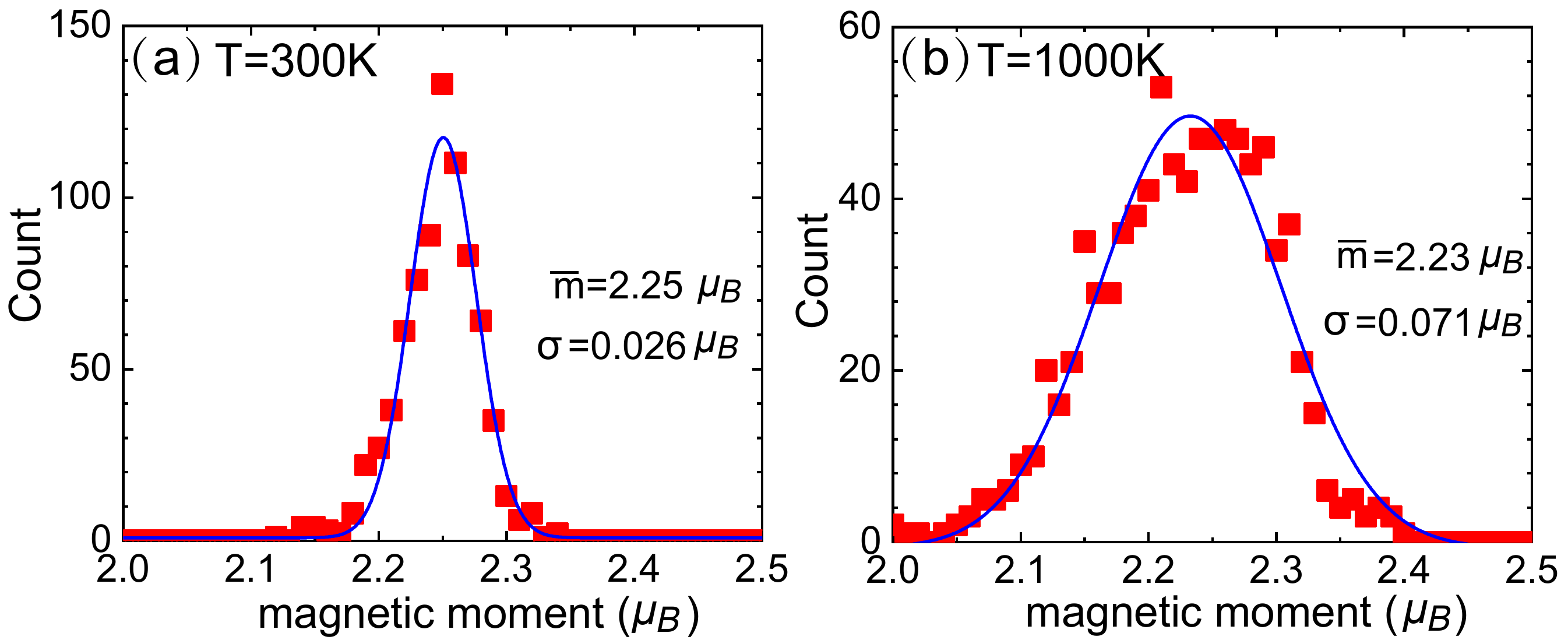}
\caption{\label{fig:mdis} Local magnetic moment distributions in the training and testing sets for (a) 300K and (b) 1000K resulting from constrained DFT calculations (red squares).
In each case the mean and standard deviations are calculated and the corresponding Gaussian distributions are superimposed (thin blue line). The moment distributions are counted using an interval of $0.01 \mu_B$.
}
\end{figure}

Though the configurations $\{{\bf e}_i^\alpha \}$ were generated using a Heisenberg model in which the Fe atomic magnetic moments were fixed in magnitude ($=2.23 \mu_{\rm B}$), these moments are free to vary in the constrained DFT calculations.  
For each temperature (300 K and 1000 K), there are 50 configurations of 16 atomic moments and the distribution of their amplitudes are plotted in \Cref{fig:mdis}. 
In each case, we calculate the mean and standard deviation and plot the corresponding Gaussian distribution (as a thin blue line). 
The mean values, 2.25$\mu_B$ for 300 K and 2.23$\mu_B$ for 1000 K are essentially the same but the standard deviation increases from $\sigma =0.026\mu_B$ to $\sigma =0.071\mu_B$ on increasing the temperature. 
At 1000K, just below the Curie temperature of bcc Fe ($T_C=1043\,$K), the variation of the magnetic moment within one standard deviation, is $\sim 0.071/2.23 = 3.2$\% while the RMSE of the predicted total energy is 23 meV ($\sim 1.44\,$meV/atom), corresponding to a $\sim 1$\% error.
This slow variation of the magnitude is implicitly learned by the ML model that does not  include $m_i$ explicitly. 

\begin{figure}[t]
\includegraphics[width=8.6cm]{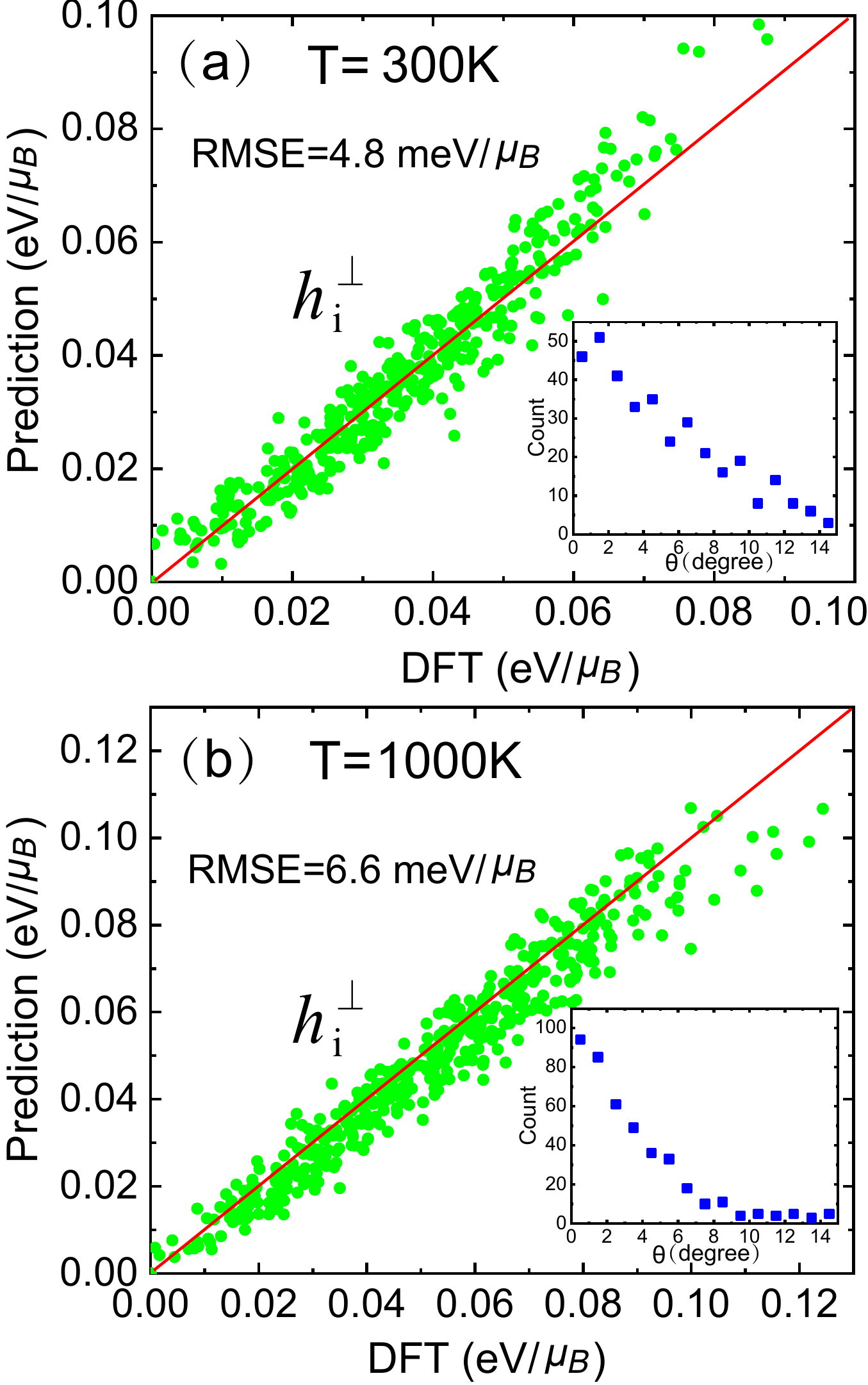}
\caption{\label{fig:DFT-B} Transverse exchange field $h_i^\perp$ predicted by the ML-FF model compared with the constrained-DFT values for (a) 300K and (b) 1000K. 
The insets show the distributions of angles between the predicted transverse exchange field and the ``exact'' ones  from constrained noncollinear DFT calculations.
}
\end{figure}
 
\subsubsection*{Exchange Fields}

The transverse component of the exchange field, ${\bf h}_i^\perp$, predicted by the ML model using \eqref{III3} with a fixed value of $m_i=2.23 \mu_B$ is shown (red arrows) in the second row of \Cref{fig:300-1000} for a single spin configuration $\{ {\bf e}_i^\alpha \}$ (grey arrows) for each temperature. 
These should be compared to the corresponding ``exact'' exchange fields calculated self consistently and shown in the bottom row. 
The results indicate that the model predicts not only the total energy but the atom-resolved exchange fields quite well. 
As noted in \Cref{ssec:Test-Heisenberg}, SD will not depend on $m_i$ in the adiabatic approximation because of the cancellation in \eqref{IIIA4}.

In quantitative terms, the RMSE of the magnitudes $h^\perp_i$ is 4.8 and 6.6 meV$/\mu_B$ for 300K and 1000K, respectively (\Cref{fig:DFT-B}), more than three times larger than what we saw for the Heisenberg model in \Cref{fig:HHm}. 
The larger RMSE predicted for the DFT data can be attributed in part to the larger value of $\sigma_m$ found with DFT and the fixed magnitude of moment (2.23$\mu_B$) used in \eqref{III3}. 
Though the orientations of ${\bf h}^\perp_i$ from the DFT calculations align reasonably well with the ML predictions, the insets to \Cref{fig:DFT-B} illustrate the limitations of the statistics available with these small supercells. 
Because of the expensive computational cost for noncollinear constrained DFT calculations, only 25 samples were used for training. 
As discussed in \Cref{sssec:TH-m} for the Heisenberg Hamiltonian, using a ten times larger training set can improve the exchange-field prediction by roughly 50\%. 
We expect to be able to reduce the RMSE of the predicted energy and exchange field by considering larger training sets and the three-body descriptor derived in \Cref{Appendix:A}. 

\subsubsection*{Portability}

\begin{figure}[t]
\includegraphics[width=8.6cm]{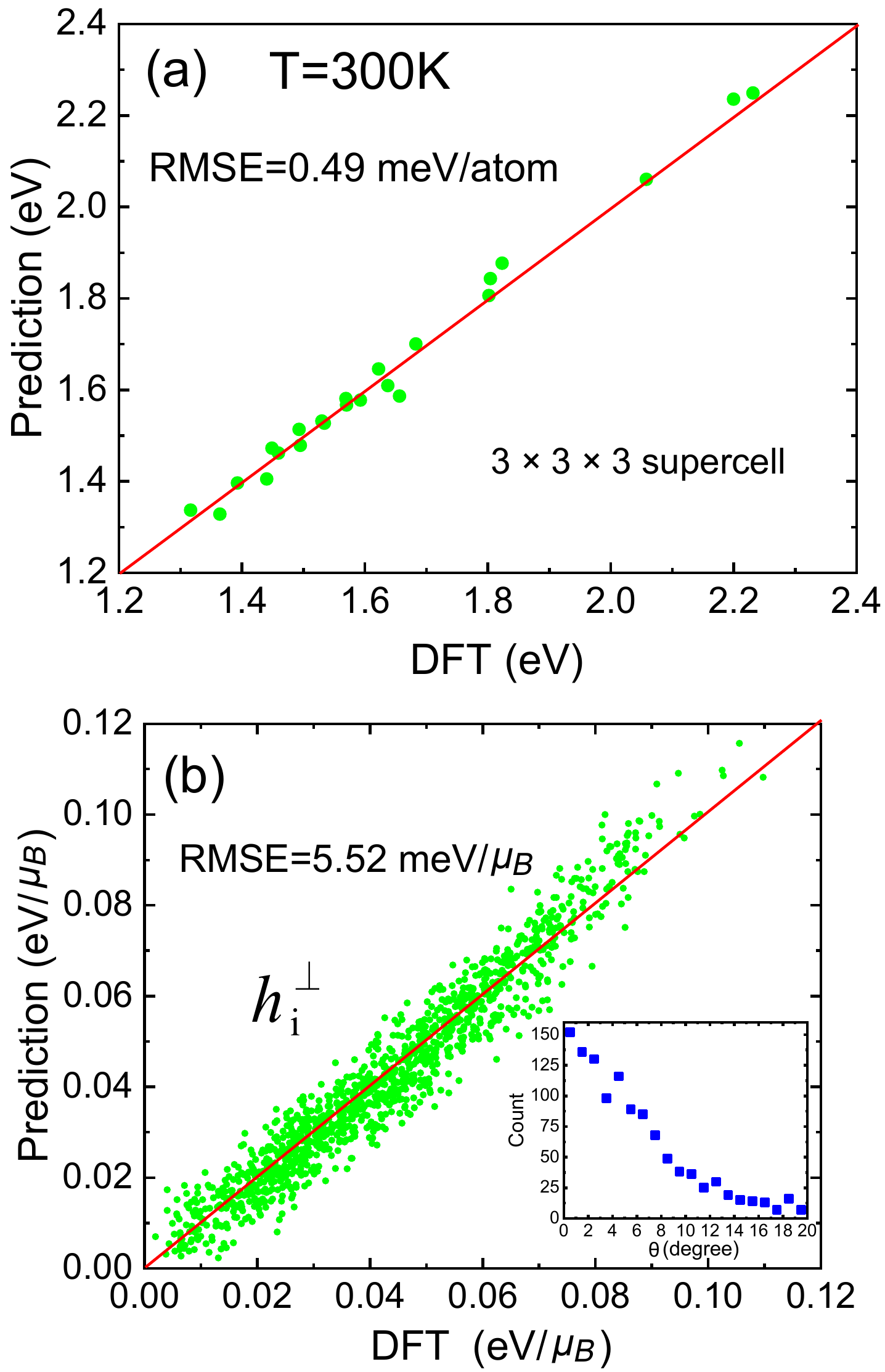}
\caption{\label{fig:54}Comparison of (a) total DFT energies and (b) transverse exchange fields calculated for 23 noncollinear spin configurations in 54-atom supercells of Fe with those predicted using the ML model at 300K.  
The ML model was trained with data calculated for a 16 atom supercell of bcc Fe. }
\end{figure}

To assess the portability of the exchange fields, we used the ML model trained with data for 16 atom supercells of bcc Fe at 300K to predict the total energies for  a 3$\times 3 \times$3 cubic supercell containing 54 Fe atoms at the same temperature. 
Because of the computational expense of constrained noncollinear calculations in larger supercells ($\sim 4000$ core hours for a single spin configuration in the 3$\times 3 \times$3 case), only 23 samples were selected for testing. 
As shown in \Cref{fig:54}, the model predicts the total energy of the larger supercell very well with an RMSE of 0.49 meV/atom that is even smaller than the 0.68 meV/atom found for the 2$ \times 2 \times $2 supercell in \Cref{fig:300-1000}(a).
The RMSE of 5.5 meV/$\mu_B$ on the transverse exchange field is slightly larger than corresponding value of 4.8 meV/$\mu_B$ shown for the 2$ \times 2 \times $2 case in \Cref{fig:DFT-B}(a).

\section{Discussion}
\label{Sec:Discussion}

A specific motivation for this work, apart from a general interest in the first-principles modelling of magnetic materials, was the need to be able to model thermal disorder realistically \cite{LiuY:prb11} in order to study the transport properties of magnetic materials, in particular novel materials \cite{Gibertini:natn19} whose temperature dependent properties are not well documented.

In Ref.~\cite{LiuY:prb15}, we attempted to establish a fully ``first-principles'' framework of finite-temperature quantum transport by basing thermal lattice and spin disorder on phonon and magnon dispersion relations calculated from first-principles. 
That approach was, however, limited to small amplitudes of atomic displacements and moment rotations, respectively. 
Limitations of the harmonic approximation for phonons can be circumvented by carrying out AIMD simulations to generate ``snapshots'' of lattice disorder within the adiabatic approximation.
For well-documented materials, it is possible to choose Gaussian spin disorder to match the experimental magnetization behaviour and Gaussian lattice disorder so that together the experimental resistivity is reproduced \cite{LiuY:prl14, WangL:prl16, Gupta:prl20, Nair:prl21}.
For novel materials whose magnetization $M(T)$ has not yet been measured, the lack of a corresponding approach to spin disorder led to the present work.
An additional aspect we wanted to be able to consider was the coupling between the spin and spatial coordinates (magnon-phonon coupling) as spin dynamics occurs on timescales ($\sim$ picoseconds) comparable to those encountered in molecular dynamics.

This motivated our decision to base the present work on plane-wave based DFT frameworks capable of AIMD. 
In the ``adiabatic spin approximation'', the descriptor only depends on the spin orientations and the atomic coordinates where the spins are located. 
Introduction of ``the smooth overlap of spin orientations'' (SOSO) allows the similarity of two local spin environments to be accurately assessed. 
A formulation of the descriptor and similarity kernel for the exchange interaction (and in \Cref{Appendix:B} for the  magnetic dipole-dipole interaction) was presented and a two-body descriptor for the exchange interaction was implemented and tested. 
This descriptor predicts the total energy and exchange field of noncollinear spin systems remarkably well even when trained with a small training set; at a level of 1~meV per spin using just 25 DFT training energies. 
We attribute this success to the physical simplicity of the dominant magnetic interactions, suggesting that the energy landscape is largely governed by the short-range Heisenberg Hamiltonian, $J(r) \, {\bf m}_i \cdot {\bf m}_j $, used ubiquitously in phenomenological studies of magnetism. 
Although both $J(r)$ and ${\bf m}_i$ may vary with local spin configurations, the relationship 
$J(r) \, {\bf m}_i \cdot {\bf m}_j 
= J(r) \, m_i m_j \, {\bf e}_i \cdot {\bf e}_j 
= \bar{J}(r) \, {\bf e}_i \cdot {\bf e}_j $ implies that the model only needs to learn an effective function $\bar{J}(r)$ for a limited set of interatomic distances. 
This constitutes a low-dimensional problem well within the capacity of our ML-EF approach.

Although the degrees of freedom corresponding to the moment magnitude $m_i$ were not explicitly included, the ML-EF model does describe a slow adiabatic variation of the magnetic moments implicitly, an aspect not considered in the original treatment of AISD \cite{Antropov:prl95, *Antropov:prb96}. 
The excellent reproduction of the DFT data demonstrates that the dependence of the exchange interaction on the spin environment is well captured by the ML-EF model.
It suggests that explicit inclusion of the moments $\{m_i\}$ is not necessary to perform an AISD simulation; the absence of $m_i$ in the equation of motion \eqref{IIIA5} in the adiabatic spin approximation reinforces this conclusion.  
Though including $\{m_i \}$ will increase the accuracy of the ML model, it comes at the cost of enlarging the dimensionality of the descriptor making training computationally more expensive. 
The adiabatic assumption relies on distinct time scales for the evolution of the spin orientation ${\bf e}_i$ and its magnitude $m_i$.
When these time scales are comparable, for example in the field of ultrafast demagnetization \cite{Kirilyuk:rmp10}, it may well become invalid. 

Though the exchange interaction is intrinsically two-body, the DFT interaction we wish to describe includes both exchange and correlation and it is not {\it a priori} clear how important three-body terms will be. 
The good results we have presented with just a two-body descriptor suggest that the latter are significantly less important. 
While the three-body terms worked out in the Appendices are expected to improve accuracy \cite{Rinaldi:npjcm24} they will require larger training sets where we are already at the limit of what is computationally possible. 

\subsubsection*{Other work}
\label{sssec:ow}

While many of the ideas developed for ML-FFs have been applied to the description of magnetic materials based upon DFT calculations \cite{Eckhoff:npjcm21, Novikov:npjcm22, Kotykhov:cms24, Suzuki:prb23, Li:ncs23, Rinaldi:npjcm24} or model Hamiltonians \cite{ZhangP:prl21, *ZhangP:npjcm23, Domina:prb22, Chapman:sr22, YuH:prb22, *YuH:prb24}, our particular interest is in DFT-based treatments of non-collinear magnetism \cite{Suzuki:prb23, Rinaldi:npjcm24}.

Suzuki's extension \cite{Suzuki:prb23} of the SNAP representation \cite{Thompson:jcompp15} to the GAP-SOAP framework is similar in spirit to the present work. 
However, where Suzuki {\it et al.} construct effective magnetization densities by replacing the delta function of point multipoles with the product of a spatial Gaussian times the multipole moments, we replace the delta-function orientation of the magnetic moment with a Gaussian which allows us to evaluate the similarity between spin configurations with smooth overlaps of both atomic positions and spin orientations.
Focussing on materials with complex magnetic anisotropies, Suzuki {\it et al.} found that a higher-order partial spectrum (trispectrum) is needed to distinguish magnetic structures with different magnetic anisotropies.
That is beyond the current state-of-the-art of DFT calculations which allows magnetic anisotropy energies (MAE) of the order of 1~meV/atom for uniaxial systems to be qualitatively determined \cite{Daalderop:prb90b, Daalderop:prl92, Daalderop:prb94, Steiner:prb16, Dieny:rmp17, Halder:prb23} but not the smaller MAEs of higher symmetry systems that are $<0.1$meV/atom \cite{Daalderop:prb90a, Daalderop:prb91, Stiles:prb01}.
Calculation of the MAE requires inclusion of the spin-orbit interaction and extremely fine k-point sampling making it much more computationally expensive to produce training sets. 
In view of the expense of the noncollinear calculations used in \Cref{ssec:Test-DFT}  without SOC, we identify the lack of a first-principles framework to calculate higher-order MAEs as a major bottleneck to developing ML-FFs for magnetic materials. 

\subsubsection*{Magnetic ACE}
\label{sssec:MLvsMCE}

The atomic cluster expansion (ACE) method is a systematic approach to a general representation of interatomic interactions in terms of body-ordered (two-body, three-body and higher) interactions \cite{Drautz:prb19}. The generalization of the formalism to spin degrees of freedom (DoF) \cite{Drautz:prb20} was very recently applied to the description of iron including noncollinear magnetism \cite{Rinaldi:npjcm24}. 
Insofar as Rinaldi {\it et al.} included the magnitude of the atomic moments, their implementation is more general than what we have presented. 
A very extensive training set containing more than 70000 structures was used including different crystal structures of Fe with a view to studying phase transitions at finite temperatures whereby noncollinearity played a minor role. 
Rather than performing simultaneous MD and SD when both spin and lattice DoF were included, the spin DoF were treated using Monte Carlo sampling because SD based on the LLG equation that only evolves the spin direction with time is not compatible with their ML model that demands that both spin direction and magnitude be able to vary.

Compared to the magnetic ACE (mACE) that includes up to fourth-order contributions \cite{Drautz:prb20, Rinaldi:npjcm24}, we showed that a descriptor for the ``two-body'' interatomic exchange interaction describes the total magnetic energy and exchange field accurately using a small training set. 
 Unlike the mACE model that includes both  magnitude and direction of the magnetic moment \cite{Rinaldi:npjcm24}, we introduced an ``adiabatic approximation'' for the magnitude of the magnetic moment and the resulting SOSO descriptor only includes the  directions and positions of magnetic moments explicitly; the intraatomic exchange interaction is accounted for implicitly leading to a high computational efficiency. 
As such, our ML model can be integrated into the equation of motion of the AISD scheme proposed by Antropov \cite{Antropov:prl95, *Antropov:prb96}.

\section{Conclusion}
\label{Sec:Conclusion}

We have derived a number of descriptors within the GAP-SOAP framework that have the rotational symmetries of specific terms in the spin Hamiltonians used to model noncollinear spin systems \eqref{Eq:SH}, in particular the interatomic exchange interaction. 
Use of the ``adiabatic spin approximation'' in the resulting ML model contributes greatly to its efficiency requiring only small constrained DFT training sets to achieve robust performance. 
For example, only 25 different noncollinear spin configurations were needed to train a simple two-body descriptor for the exchange interaction of bcc Fe that could predict the total energy with an accuracy of order 1 meV/spin and of the exchange field with an angular deviation less than $\sim5^\circ$. 
By incorporating atomic coordinates and spin orientations into the local spin environment descriptor, our model paves the way for the construction of a general ML PES that will make coupled ab-initio molecular dynamics (AIMD) and spin dynamics (AISD) simulations possible to study the effects of spin-lattice coupling at finite temperatures. 

\begin{acknowledgments}
This work was financially supported by the National Natural Science Foundation of China (nos.12404255) and the ``Nederlandse Organisatie voor Wetenschappelijk Onderzoek'' (NWO) through the research programme of the former ``Stichting voor Fundamenteel Onderzoek der Materie,'' (NWO-I, formerly FOM) and through the use of supercomputer facilities of NWO ``Exacte Wetenschappen'' (Physical Sciences). 
The work described in this manuscript required implementing our GAP-SOSO framework into the {\sc quip} code \cite{Csanyi:07, Kermode:jpcm20, Bartok:prl10} available under a non-commercial academic source license and the {\sc ml-ef} source code and data presented in the manuscript are available in \cite{ML-EF:GitHub}.  
\end{acknowledgments}

\section*{DATA AVAILABILITY}
The data that support the findings of this article are openly available at \cite{ML-EF:GitHub}.

\appendix

\begin{widetext}
\section{A three-body descriptor for the interatomic exchange interaction. }
\label{Appendix:A}

The exchange interaction between two spins includes a part that depends on all of the other spins \cite{Heine:prb21, Szilva:prb17}. 
Motivated by the ``angular'' three-body term in ML-FFs \cite{Behler:prl07, Bartok:prb13, Jinnouchi:prb19}, we consider a contribution to the exchange interaction that depends on the three vectors ${\bf m}_i, {\bf m}_j$ and ${\bf m}_k$ as well as the separations $d_{ij}$ and $d_{ik}$ from ${\bf m}_i$, where the Hamiltonian is invariant under simultaneous rotations of ${\bf m}_i, {\bf m}_j$ and ${\bf m}_k$. 
By analogy with \eqref{IIC1}, we consider the distribution function 
\begin{equation}
\label{A1}
\rho_i({\bf r}_1,{\bf r}_2,{\bf e}_1,{\bf e}_2,{\bf e}_3)= \sum_{j \neq k}^{N_a} f_{\rm cut}(r_{ij}) g({\bf r}_1-{\bf r}_{ij})   
 g({\bf r}_2-{\bf r}_{ik}) g({\bf e}_1 - {\bf e}_i) g({\bf e}_2 - {\bf e}_j)  g({\bf e}_3 - {\bf e}_k) 
\end{equation}
where integrating over the angular parts of ${\bf r}_1$ and ${\bf r}_2$ allows us to define the analogue of \eqref{IIC2}
\begin{equation}
\label{A2}
\rho_i(r_1,r_2,{\bf e}_1,{\bf e}_2,{\bf e}_3) = 
\iint d{\bf \hat{r}}_1 d{\bf \hat{r}}_2 
           \rho_i({\bf r}_1,{\bf r}_2,{\bf e}_1,{\bf e}_2,{\bf e}_3).
\end{equation}
The similarity $S$ of two such distributions $\rho_i$ and $\rho'_i$ is 
%
\begin{subequations}
\begin{align}
\label{A3}
 S(\widehat{R})=\iiint r_1^2 dr_1 \, r_2^2 dr_2 \, d{\bf e}_1 \, d{\bf e}_2 \, d{\bf e}_3 \,\rho_i(r_1,r_2,{\bf e}_1,{\bf e}_2,{\bf e}_3)
  \rho'_i(r_1,r_2,\widehat{R}{\bf e}_1,\widehat{R}{\bf e}_2,\widehat{R}{\bf e}_3) \\
= \sum_{\substack {n_1,n_2,m,m_1,m_2 \\
                   m',m'_1,m'_2 \\
                   l,l_1,l_2}} C_{n_1n_2ll_1l_2mm_1m_2}^* C'_{n_1n_2ll_1l_2m'm'_1m'_2} 
    D_{mm'}^{l}(\widehat{R})D_{m_1m'_1}^{l_1} (\widehat{R})D_{m_2m'_2}^{l_2}(\widehat{R})
\end{align}
\end{subequations}
where $C_{n_1n_2ll_1l_2mm_1m_2}=\frac{1}{\sqrt{4\pi}}c_{n_1} c_{n_2} C_{lm} C_{l_1m_1} C_{l_2m_2}$ and we drop the atom-spin index $i$ on the coefficients to avoid overloading the notation. 
$c_{nlm}$ was defined by \eqref{IIA7} and $c_n \equiv c_{nlm}$ by \eqref{IIA9} so that we can write
\begin{gather}
\label{A4} 
\rho_i(r_1,r_2) =\iint d{\bf \hat{r}}_1 d{\bf \hat{r}}_2 \sum_{j \neq k}^{N_a} f_{\rm cut}(r_{ij}) f_{\rm cut}(r_{ik})g({\bf r}_1-{\bf r}_{ij})   
 g({\bf r}_2-{\bf r}_{ik})
 =\frac{1}{4\pi} \sum_{n_1=1}^{N_R}c_{n_1} \chi_{n_1l}(r_1) \sum_{n_2=1}^{N_R}c_{n_2} \chi_{n_2l}(r_2)
\end{gather}
in terms of the orthonormal radial basis $\chi_{nl}(r)$. 

By making use of the important property of the Wigner  matrices which follows from the commutation of rotation and time reversal operators \cite{Wigner:59},
\begin{equation}
\label{A5}
D_{mm'}^{l}(\widehat{R})=(-1)^{m-m'} D_{-m,-m'}^{l^{\,*}}(\widehat{R})
\end{equation}
the integral of three $D$ matrices can be expressed as  
\begin{subequations}
\label{A6}
\begin{align}
\int D_{mm'}^{l}(\widehat{R}) 
D_{m_1m'_1}^{l_1}(\widehat{R}) 
D_{m_2m'_2}^{l_2}(\widehat{R}) d\widehat{R}   
&=(-1)^{m-m'}\int D_{-m,-m'}^{l^{\,*}}(\widehat{R}) 
D_{m_1m'_1}^{l_1}(\widehat{R}) 
D_{m_2m'_2}^{l_2}(\widehat{R}) d\widehat{R}  \\
&=(-1)^{m-m'}\frac{8\pi^2}{2l+1} \langle l,-m|l_1m_1;l_2m_2 \rangle \langle l,-m'|l_1m'_1;l_2m'_2 \rangle
\end{align}
\end{subequations}
in which the angle brackets are Clebsch–Gordan coefficients which are only nonzero when
\begin{subequations}
\label{A7}
\begin{align}
|l_1-l_2| \le \; &l \le l_1+l_2 \\
m_1+m_2=-m  \;\;\; &; \;\;\; m'_1+m'_2=-m'. 
\end{align}
\end{subequations}
The rotationally invariant similarity kernel for the three-body descriptor becomes
\begin{subequations}
\label{A8}
\begin{align}
 k(\rho_i,\rho'_i) &=\int S(\widehat{R})d\widehat{R} 
= \sum_{\substack {m,m_1,m_2 \\
                   m',m'_1,m'_2 \\
                   l,l_1,l_2,n_1,n_2}} C_{n_1n_2ll_1l_2mm_1m_2}^* C'_{n_1n_2ll_1l_2m'm'_1m'_2} 
  \int D_{mm'}^l(\widehat{R}) D_{m_1m'_1}^{l_1} (\widehat{R})D_{m_2m'_2}^{l_2}(\widehat{R}) d\widehat{R}  \label{A8a} \\ 
  = &\sum_{\substack {m,m_1,m_2 \\
                      m',m'_1,m'_2 \\
                      l,l_1,l_2,n_1,n_2}} 
                      C_{n_1n_2ll_1l_2mm_1m_2}^* C'_{n_1n_2ll_1l_2m'm'_1m'_2} 
 (-1)^{m-m'} \int D^{l^{\,*}}_{-m,-m'} (\widehat{R}) D_{m_1m'_1}^{l_1} (\widehat{R})D_{m_2m'_2}^{l_2}(\widehat{R}) d\widehat{R}  \label{A8b} \\ 
  = &\sum_{\substack {m,m_1,m_2 \\
                      m',m'_1,m'_2 \\
                      l,l_1,l_2,n_1,n_2}} C_{n_1n_2ll_1l_2mm_1m_2}^* C'_{n_1n_2ll_1l_2m'm'_1m'_2} \frac{8\pi^2(-1)^{m+m'}}{2l+1}  
   \langle l,-m|l_1m_1;l_2m_2 \rangle \langle l,-m'|l_1m'_1;l_2m'_2 \rangle \label{A8c} 
\end{align}
\end{subequations}
with
\begin{equation}
\label{A9}
C_{n_1n_2ll_1l_2mm_1m_2}
=\frac{1}{4\pi} c_{n_1} c_{n_2}  \sum_{j,k}^{N_a} \Bigg(\frac{4}{I_0(\cfrac{1}{\sigma^2})}\Bigg)^3  
\iota_l \Big(\frac{1}{\sigma_s^2}\Big)
\iota_{l_1} \Big(\frac{1}{\sigma_s^2}\Big)   
\iota_{l_2} \Big(\frac{1}{\sigma_s^2}\Big) 
 Y_{lm}^*({\bf e}_i) 
 Y_{l_1m_1}^*({\bf e}_j)
 Y_{l_2m_2}^*({\bf e}_k) .
\end{equation}
Then, making the substitution 
\begin{equation}
\label{A10}
\widetilde C_{n_1n_2ll_1l_2,mm_1m_2}
= C_{n_1n_2ll_1l_2mm_1m_2} (-1)^m \sqrt{\frac{8\pi^2}{2l+1}} 
                          \langle l,-m|l_1m_1;l_2m_2 \rangle 
\end{equation}
it can finally be written in the form 
\begin{equation}
\label{A11}
k(\rho_i,\rho'_i)= \sum_{\substack {m,m_1,m_2 \\
                      m',m'_1,m'_2 \\
                      l,l_1,l_2,n_1,n_2}}  
    \widetilde C_{n_1n_2ll_1l_2,mm_1m_2}^*  \widetilde{C}'_{n_1n_2ll_1l_2,m'm'_1m'_2}.
\end{equation} 

Depending on the symmetry of the magnetic interaction we want to describe, we can define other three-body descriptors by imposing different constraints. 
\eqref{A9} was derived by requiring the invariance of three spin orientations under a rotation $\widehat{R} \equiv \widehat{R}_{\bf e}$ in spin space.
If we require invariance under separate rotations of the relative atomic positions ($r_{ij}$, $r_{ik}$) and of the spin orientations (${\bf e}_i$, ${\bf e}_j$, ${\bf e}_k$), we derive a three-body descriptor that is intermediate between those describing the magnetic dipole-dipole (\Cref{Appendix:B}) and exchange interactions. 
This three-body term may describe noncollinearity arising from the interaction between neighbouring spins more efficiently. 

In the absence of spin-orbit coupling, we consider separate rotations $\widehat{R}_{\bf r}$ of the atomic positions and $\widehat{R}_{\bf e}$ of the spin orientations. Starting with the distribution function \eqref{A1}, the similarity becomes
\begin{subequations}
\label{A12}
\begin{align}
S(\widehat{R}_{\bf r},\widehat{R}_{\bf e}) &= \iiint r_1^2 dr_1 r_2^2 dr_2 
  d{\bf e}_1 d{\bf e}_2 d{\bf e}_3 \,
  \rho({\bf r}_1,{\bf r}_2,{\bf e}_1,{\bf e}_2,{\bf e}_3) \,
  \rho'(\widehat{R}_{\bf r} {\bf r}_1, \widehat{R}_{\bf r} {\bf r}_2, \widehat{R}_{\bf e} {\bf e}_1, \widehat{R}_{\bf e} {\bf e}_2, \widehat{R}_{\bf e} {\bf e}_3) \label{A12a} \\ 
&  = \sum_{\substack {n_1,n_2,m,m_1,m_2 \\
                     m',m'_1,m'_2 \\
                     M_1,M_2,M'_1,M'_2 \\
                     l,l_1,l_2,L_1,L_2 }} 
 B_{n_1n_2L_1L_2M_1M_2}^* B'_{n_1n_2L_1L_2M'_1M'_2} 
 C_{l{l_1}{l_2}mm_1m_2}^* C'_{ll_1l_2m'm'_1m'_2} \nonumber \\
 & \;\;\;\;\; \;\;\;\;\; \;\;\;\;\; \;\;\;\;\; \;\;\;\;\;  \;\;\;\;\;\times D_{M_1M'_1}^{L_1}(\widehat{R}_{\bf r})  D_{M_2M'_2}^{L_2} (\widehat{R}_{\bf r}) 
  D_{mm'}^{l}(\widehat{R}_{\bf e}) D_{m_1m'_1}^{l_1} (\widehat{R}_{\bf e}) 
  D_{m_2m'_2}^{l_2}(\widehat{R}_{\bf e})   \label{A12b}
\end{align}
\end{subequations}
and the corresponding kernel for the three-body descriptor can be written
\begin{subequations}
\label{A13}
\begin{align}
\!\!\! \!\!\! \!\!\! \!\!\! k(\rho_i,\rho'_i) &=  
\iint S(\widehat{R}_{\bf r},\widehat{R}_{\bf e}) d\widehat{R}_{\bf r}d\widehat{R}_{\bf e} 
=  \!\!\! \!\!\! \!\!\! \!\!\! \sum_{\substack {n_1, n_2, l, l_1, l_2 \\
                   m,m_1,m_2, m',m'_1, m'_2 \\
                   L_1, L_2, M_1,M_2,M'_1,M'_2}}  \!\!\! \!\!\! \!\!\! \!\!\!
B_{n_1n_2L_1L_2M_1M_2}^* B'_{n_1n_2L_1L_2M'_1M'_2} 
    C_{ll_1l_2mm_1m_2}^* C'_{ll_1l_2m'm'_1m'_2}  \nonumber \\
& \;\;\;\;\;  \;\;\;\;\;  \;\;\;\;\;  \;\;\;\;\;  \;\;\;\;\; 
  \times \int D_{M_1M'_1}^{L_1}(\widehat{R}_{\bf r}) D_{M_2M'_2}^{L_2} (\widehat{R}_{\bf r}) \, d\widehat{R}_{\bf r}  
  \int D_{mm'}^l(\widehat{R}_{\bf e}) D_{m_1m'_1}^{l_1} (\widehat{R}_{\bf e})D_{m_2m'_2}^{l_2}(\widehat{R}_{\bf e}) \, d\widehat{R}_{\bf e}  \label{A13a} \\
= \sum_{\substack {n_1, L_1, M_1, M'_1 \\
                   n_2, L_2, M_2, M'_2 }} 
  &  B_{n_1n_2L_1L_2M_1M_2}^* B'_{n_1n_2L_1L_2M'_1M'_2} 
 (-1)^{M_1-M'_1} \int D^{L_1^*}_{-M_1,-M'_1} (\widehat{R}_{\bf r}) 
                  D^{L_2}_{M_2,M'_2} (\widehat{R}_{\bf r}) 
                                     d\widehat{R}_{\bf r} \nonumber \\
               \times & \sum_{\substack {l, l_1, l_2 \\
                                         m, m_1, m_2 \\
                                         m',m'_1,m'_2}}
  C_{ll_1l_2mm_1m_2}^* C'_{ll_1l_2m'm'_1m'_2} (-1)^{m-m'}               
 \int D^{l^*}_{-m,-m'}(\widehat{R}_{\bf e}) 
      D^{l_1}_{m_1m'_1}(\widehat{R}_{\bf e}) 
      D^{l_2}_{m_2m'_2}(\widehat{R}_{\bf e}) d\widehat{R}_{\bf e} \label{A13b} \\  
 =\sum_{\substack {n_1, n_2, L_1, M_1, M'_1}}  & \frac{8\pi^2(-1)^{M_1+M'_1}}{2L_1+1}  B_{n_1 n_2 L_1 L_1 M_1, -M_1}^* B'_{n_1n_2L_1L_1M'_1,-M'_1} 
 \sum_{\substack {l, l_1, l_2 \\
                  m, m_1, m_2 \\
                  m',m'_1,m'_2}}
                   C_{ll_1l_2mm_1m_2}^* C'_{ll_1l_2m'm'_1m'_2} \nonumber \\    
 & \;\;\;\;\;\;\;\;\;\;\;\;\;\;\;\;\;\;\;\;\;\;\;\;\;\;\;\;\;\;\;\; \times \frac{8\pi^2(-1)^{m+m'}}{2l+1} \langle l,-m|l_1m_1;l_2m_2 \rangle \langle l,-m'|l_1m'_1;l_2m'_2 \rangle  \label{A13c}  \\ 
= \sum_{\substack {n_1,n_2,L_1}} &
P_{n_1n_2L_1} P'_{n_1n_2L_1}  
\sum_{\substack {l, l_1, l_2 \\
                 m, m_1, m_2 \\
                 m',m'_1,m'_2}} C_{ll_1l_2mm_1m_2}^* C'_{ll_1l_2m'm'_1m'_2} \nonumber \\
& \;\;\;\;\;\;\;\;\;\;\;\;\;\;\;\;\;\;\;\;\;\;\;\;\;\;\;\;\;\;\;\; \times \frac{8\pi^2(-1)^{m+m'}}{2l+1}  \langle l,-m|l_1m_1;l_2m_2 \rangle  
                                \langle l,-m'|l_1m'_1;l_2m'_2 \rangle  \label{A13d} 
\end{align}
\end{subequations} 
where   
\begin{equation}
\label{A14} 
C_{ll_1l_2mm_1m_2}= \sum_{j,k}^{N_a} \Bigg(\frac{4}{I_0(\cfrac{1}{\sigma^2})}\Bigg)^3  
\iota_l \Big(\frac{1}{\sigma_s^2}\Big)
\iota_{l_1} \Big(\frac{1}{\sigma_s^2}\Big)   
\iota_{l_2} \Big(\frac{1}{\sigma_s^2}\Big) 
 Y_{lm}^*({\bf e}_i) 
 Y_{l_1m_1}^*({\bf e}_j)
 Y_{l_2m_2}^*({\bf e}_k) ,
 \end{equation}
\begin{equation}
\label{A15} 
B_{n_1 n_2 L_1 L_2 M_1 M_2}= c_{n_1 L_1 M_1} \; c^*_{n_2 L_2 M_2} ,
\end{equation}
and 
\begin{equation}
\label{A16} 
P_{n_1n_2L_1}=\sqrt{\frac{8\pi^2}{2L_1+1}}
               \sum_{M=-L_1}^{L_1}(-1)^M B_{n_1n_2L_1L_1M_1,-M_1}^*
 = \sqrt{\frac{8\pi^2}{2L_1+1}}\sum_{M=-L_1}^{L_1} c^*_{n_1 L_1 M_1} c_{n_2 L_1 M_1}
\end{equation}
is nothing other that the power spectrum defined in \eqref{IIA11} \cite{Bartok:prb13}.

\section{Descriptor for magnetic dipole-dipole like interaction}
\label{Appendix:B}

When performing MD simulations with ML-FFs for polar solids, a standard partitioning scheme is used to calculate equivalent multipoles of the local charge density distribution $\rho_q({\bf r})$. 
When these are subtracted from $\rho_q({\bf r})$, the multipoles of the resulting charge density distribution $\tilde{\rho}_q({\bf r})$  are identically zero so the electrostatic interactions involving $\tilde{\rho}_q({\bf r})$ have short range and can be calculated in reciprocal space as part of the quantum mechanical calculation  \cite{[An illustration of the procedure can be found in Appendix A of ] Kittel:56}.
The long-range interactions of $\rho_q({\bf r})$ are calculated ``classically'' using the equivalent point multipoles as part of the conventional MD simulation. 

Magnetic multipolar interactions can be handled analogously and we derive a descriptor to handle the deviation of magnetic multipoles from their point form; because of the weakness of the magnetic dipole-dipole interactions, we expect these short range corrections to be very small.
Unlike the interatomic exchange interaction, the classical magnetic dipole-dipole interaction in \eqref{Eq:SH} depends on ${\bf m}_i$ and ${\bf m}_j$ as well as on their relative position ${\bf r}_{ij}$, \Cref{fig:spin-model}. 
${\bf m}_i$, ${\bf m}_j,$ and ${\bf r}_{ij}$ should be coupled in the descriptor to keep the rotational symmetry of the dipole-dipole interaction. 
In \eqref{Eq:SH}, the magnetic exchange and dipole-dipole interactions have two-body form but with different rotational symmetries. 
Here, we suggest a descriptor for magnetic dipole-dipole-like interactions that maintains the rotational symmetry of ${\bf e}_i, {\bf e}_j,$ and ${\bf r}_{ij}$. 
Starting from the distribution function 
\begin{equation}
\label{B1}
\rho_i({\bf r},{\bf e}_1,{\bf e}_2)
= \sum_{j=1}^{N_a} f_{\rm cut}( r_{ij}) 
g({\bf r}-{\bf r}_{ij}) 
g({\bf e}_1 - {\bf e}_i) 
g({\bf e}_2 - {\bf e}_j),
\end{equation}
the overlap between two local spin environments $\rho_i$ (unrotated) and $\rho'_i$ (rotated) is
\begin{equation}
\begin{aligned}
\label{B2}
 S(\widehat{R})&=\iiint r^2 d{\bf \hat{r}} \, d{\bf e}_1 \, d{\bf e}_2 
 \, \rho_i({\bf r},{\bf e}_1,{\bf e}_2) \,
 \rho'_i(\widehat{R}{\bf r}, \widehat{R}{\bf e}_1,\widehat{R}{\bf e}_2) \\
   &= \sum_{\substack {n, l,   m,   m'\\
                          l_1, m_1 ,m'_1 \\
                          l_2, m_2, m'_2 }}
    A_{nll_1l_2;mm_1m_2}^* A'_{nll_1l_2;m'm'_1m'_2} 
      D_{mm'}^{l}({\widehat{R}}) 
      D_{m_1m'_1}^{l_1}(\widehat{R}) 
      D_{m_2m'_2}^{l_2}(\widehat{R}).
\end{aligned}
\end{equation}
where $A_{nll_1l_2;mm_1m_2}=\frac{1}{\sqrt{4\pi}} c_{nlm} C_{l_1m_1} C_{l_2m_2}$ in terms of $c_{nlm}$ \eqref{IIA7} and $C_{lm}$ \eqref{IIB6} and we have omitted the atom index $i$ in the expansion coefficients. 
The similarity kernel can be written as
\begin{subequations}
\label{B3}
\begin{align}
k(\rho_i,\rho'_i) & =\int S(\widehat{R})d\widehat{R}  = 
\sum_{\substack {n, l,   m,   m'\\
                    l_1, m_1 ,m'_1 \\
                    l_2, m_2, m'_2 }}
                 A_{nll_1l_2mm_1m_2}^* A'_{nll_1l_2m'm'_1m'_2} 
      \int D_{mm'}^l(\widehat{R}) 
           D_{m_1m'_1}^{l_1} (\widehat{R})
           D_{m_2m'_2}^{l_2}(\widehat{R}) d\widehat{R}  \label{B3a} \\
& = \sum_{\substack {n, l,   m,   m'\\
                        l_1, m_1 ,m'_1 \\
                        l_2, m_2, m'_2 }}
                      A_{nll_1l_2mm_1m_2}^* A'_{nll_1l_2m'm'_1m'_2} 
 (-1)^{m-m'} \int D_{-m,-m'}^{l^{\,*}}(\widehat{R}) 
                  D_{m_1m'_1}^{l_1} (\widehat{R})
                  D_{m_2m'_2}^{l_2}(\widehat{R}) d\widehat{R} \label{B3b} \\
& = \sum_{\substack {n, l,   m,   m'\\
                        l_1, m_1 ,m'_1 \\
                        l_2, m_2, m'_2 }}
    A_{nll_1l_2mm_1m_2}^* A'_{nll_1l_2m'm'_1m'_2} \frac{8\pi^2(-1)^{m+m'}}{2l+1} 
   \langle l,-m|l_1m_1;l_2m_2 \rangle \times \langle l,-m'|l_1m'_1;l_2m'_2 \rangle. \label{B3c} \\
 & = \sum_{\substack {n, l,   m,   m'\\
                         l_1, m_1 ,m'_1 \\
                         l_2, m_2, m'_2 }}
  \widetilde A_{nll_1l_2mm_1m_2}^* \widetilde A'_{nll_1l_2m'm'_1m'_2} . \label{B3d}       
\end{align}
\end{subequations}
in terms of 
\begin{equation} 
\label{B4}
\widetilde A_{nll_1l_2mm_1m_2} = c^*_{nlm} \sum_{j,k;j \neq k}^{N_a}  
     \Bigg(\frac{4}{I_0(\cfrac{1}{\sigma^2})}\Bigg)^2 
        \iota_{l_1}\Big(\frac{1}{\sigma_s^2}\Big)  \iota_{l_2}\Big(\frac{1}{\sigma_s^2}\Big) 
  Y_{l_1m_1}^*({\bf e}_j)Y_{l_2m_2}^*({\bf e}_k)  
  \sqrt{\frac{8\pi^2}{2l+1}} \langle l,-m|l_1m_1;l_2m_2 \rangle.
\end{equation}
\end{widetext}


\begin{thebibliography}{111}
\bibitem [{foo({\natexlab{a}})}]{footnote4}%
  \BibitemOpen
  \href@noop {} \bibinfo {note} {{Because the LDA is known to underestimate the equilibrium lattice constant of the ferromagnetic 3$d$ metals by about 3\%, we used the Generalized Gradient Approximation (GGA) of Perdew, Burke and Enzerhof \cite{Perdew:prl96} to describe the effects of exchange and correlation. The calculated results were obtained with a plane-wave basis and the projector augmented wave (PAW) formalism \cite{Blochl:prb94b} as implemented in the Vienna Ab-initio Simulation Package ({\sc vasp}) \cite{Kresse:prb93a, Kresse:prb96, Kresse:prb99} using an energy cut-off of 500 eV and a k-point sampling of 36$\times$36$\times$36 for a 1$\times$1$\times$1 unit cell (or its equivalent for larger unit cells) converging the energy to $10^{-6}\,$eV}}\BibitemShut {NoStop}%
\bibitem [{foo({\natexlab{c}})}]{footnote1}%
  \BibitemOpen
  \href@noop {} \bibinfo {note} {{In the {\sc quip} code from Cambridge University, the radial parts are expanded in a set of equally spaced Gaussian functions while in the {\sc vasp} ML package the radial basis functions are normalised spherical Bessel functions. The Gaussian basis functions need to be orthogonalised. The normalised spherical Bessel functions are mutually orthogonal }}\BibitemShut {NoStop}%
\bibitem [{foo({\natexlab{d}})}]{footnote10}%
  \BibitemOpen
  \href@noop {} \bibinfo {note} {{For example, if $N$ is initially set to 20 and then incremented by 10, the root-mean-square error changes by a certain amount. At some point, increasing $N$ further does not lead to any improvement. Proceeding in this way, $N$ is set to 80 to train the Heisenberg model and 40 for DFT }}\BibitemShut {NoStop}%
\bibitem [{foo({\natexlab{e}})}]{footnote5}%
  \BibitemOpen
  \href@noop {} \bibinfo {note} {{Because we approach ML-EF from the DFT point of view, we switch from the conventional potential energy notation $U$ of classical molecular dynamics to the total (electronic) energy $E$ of DFT }}\BibitemShut {NoStop}%
\bibitem [{foo({\natexlab{f}})}]{footnote8}%
  \BibitemOpen
  \href@noop {} \bibinfo {note} {{Though the onset of the equipartition regime is strictly speaking at the Debye temperature $\Theta_D$, ``in practice the high-$T$ limit is identified as the equipartition regime with linear-in-$T$ resistivity ... but can extend practically to about $T>\Theta_D/3$ in Pt, Cu and Au and to about $T>\Theta_D/5$ in Al'' \cite{Xiao:prb19}. For typical transition-metal Debye temperatures, this means for $T > 100 \,$K \cite{Nair:prl21} }}\BibitemShut {NoStop}%
\end{thebibliography}

%

\end{document}